\documentclass{article}%
\usepackage{amssymb}
\usepackage{amsmath}
\usepackage{amsfonts}
\usepackage{graphicx}
\usepackage{texdraw}
\usepackage{framed}
\usepackage[cmtip,arrow]{xy}
\usepackage{pb-diagram,pb-xy}%
\providecommand{\U}[1]{\protect\rule{.1in}{.1in}}
\input{diagrams.tex}
\begin{document}

\title{Integrable Systems and Poisson-Lie T-duality: a finite dimensional example.}
\author{S. Capriotti, H. Montani\\Centro At\'{o}mico Bariloche and Instituto Balseiro\\(8400) S. C. de Bariloche, R\'{\i}o Negro, Argentina}
\maketitle

\begin{abstract}
We study the deep connection between integrable models and Poisson-Lie
T-duality working on a finite dimensional example constructed on $SL\left(
2,\mathbb{C}\right)  $ and its Iwasawa factors $SU\left(  2\right)  $ and $B$.
We shown the way in which Adler-Kostant-Symes theory and collective dynamics
combine to solve the equivalent systems from solving the factorization problem
of an exponential curve in $SL\left(  2,\mathbb{C}\right)  $. It is shown that
the Toda system embraces the dynamics of the systems on $SU\left(  2\right)  $
and $B$.

\end{abstract}

%\tableofcontents

%\tableofcontents


\textbf{Subject classification:} symplectic geometry, classical integrable systems

\textbf{Keywords:} Poisson Lie T-duality; Dressing actions; Momentum maps;
Collective Hamiltonians
%\email{montani@cab.cnea.gov.ar}
\newpage

\section{Introduction}

The search for dualities in theoretical physics is motivated by the hope of
finding a couple of related theories in which one of them is, in some sense,
easily solved and the solutions to the second one is attained from the
solution of the former system. \emph{Poisson-Lie T-duality} is a nice example
in this direction: it is built on phase spaces having a rich structure
entailing a close connection with integrable model, exploiting the inherent
self dual character of Poisson-Lie groups in order to relate a couple of sigma
models having targets on the factors a Drinfeld double Lie group $\cite{KS-1}%
$. In references $\cite{CM}$ and $\cite{CMZ}$, PL T-duality was accurately
encoded in a hamiltonian scheme ruled by some hamiltonian actions of the
double Lie group $G$ on the cotangent bundle of its factors, where T-duality
transformations are provided by the associated momentum maps targeting on the
same coadjoint orbit. Moreover, it was realized that collective dynamics on
these hamiltonian $G$-spaces underpins the dynamic correspondence between
these models. In those references, $G$ was taken as centrally extended
Drinfeld double of a loop group and T-duality comes to relate sigma models
built on each factor of it. This scheme reveals also the role played by a WZNW
model whose reduced phase space, the shared coadjoint orbit, embraces the
dynamics of both sigma models. In all these systems, compatible dynamics are
ruled by \emph{collective hamiltonians}. Thus, the natural setting is infinite
dimensional: it is provided by phase spaces modelled on cotangent bundles of
loop groups, and the momentum maps are associated with the centrally extended
action of the double group. In spite of this, the essential issues of
T-duality can be clearly sketched in a finite dimensional context, avoiding
the specific difficulties of the infinite dimensional case.

The current work is aimed to stress the intrinsic connection of the
Poisson-Lie T-duality with integrable systems, working in a finite dimensional
framework, allowing us to concentrate on the structural facts behind this
connection. We describe the geometric structure underlying the hamiltonian
version of this duality, following references $\cite{CM}$ and $\cite{CMZ}$, by
considering a complex Lie group $G$ and its Iwasawa decomposition in the
compact factor $K$ and the soluble one, $B$. As an alternative to the standard
scheme built on hamiltonian $G$-spaces, we introduce a wider version of
T-duality in order to include schemes based on the hamiltonian action of the
Iwasawa factors, giving rise to duality classes of hamiltonian $K$ or $B$
spaces. This leads straightforwardly to the \emph{Adler-Kostant-Symes (AKS)}
theory for integrable systems $\cite{AKS}$, through the introduction of
collective dynamics. An explicit example is constructed in full detail working
on $SL\left(  2,\mathbb{C}\right)  $ and its factors in the Iwasawa
decomposition, namely $SL\left(  2,\mathbb{C}\right)  \cong SU\left(
2\right)  \times B$, involving three hamiltonian $B$-spaces: $T^{\ast
}SU\left(  2\right)  $, $T^{\ast}B$ and $\mathbb{R}^{2}$. The respective
dually related dynamical systems are a dressing invariant system in $SU\left(
2\right)  $, a kind of generalized top on $B$, and a Toda model on
$\mathbb{R}^{2}$. This last system plays an analogous role to that played by
the WZNW in loop group case, embracing the dynamics of the other systems.
Then, we use the \emph{AKS} theory to show explicitly the integrability of
these systems constructing the solution in each case, and providing a precise
meaning for the Poisson Lie T-duality transformations. By passing to the
Lagrangian framework, we show the equivalence between systems described by
bilinear forms on the corresponding tangent bundles, so that the constructed
duality relates different \emph{targets gemetries}.

It is important to point out that most of the results can be translated, with
some cares, to the infinite dimensional case and the underlying structure
works in any case. Whatever the case, we can consider the finite dimensional
case as a restriction of the loop group one to the constant map from $S^{1}$
to a Lie group.

This work is organized as follows: in Section 2 we give a description of the
geometric setting for the hamiltonian approach to PL T-duality and its
relation with the theory of integrable models, in particular with the AKS
theory. In Section 3, we describe the main features related to Iwasawa
decomposition and coadjoint orbits; in Section 4 the involved phase spaces are
presented, describing its symmetry properties; the T-duality scheme is
described in Section 5; in Section 6 we apply explicitly the AKS Theory to
solve the systems, and in Section 7 the compatible dynamics is analyzed from
Hamiltonian and Lagrangian point of view. Finally, the conclusions are
included in Section 8. 

\section{Geometric setting for Poisson Lie $T$-duality}

The standard hamiltonian approach to PL T-duality, as introduced in
$\cite{CM}$ and $\cite{CMZ}$, considers a Lie group $G$ which can be written
as a product of two subgroups $K$ and $B$, so that all of them are endowed
with a Poisson-Lie structure and their Lie algebras $\mathfrak{g}%
,\mathfrak{k},\mathfrak{b}$, such that $\mathfrak{g}=\mathfrak{k}%
\oplus\mathfrak{b}$, turn in Lie bialgebras. Hence, the PL T-duality is built
up on hamiltonian $G$-spaces: the group $G$ acts on the cotangent bundle of
its factors, giving rise to momentum maps with non trivial intersections in
$\mathfrak{g}^{\ast}$. In the loop group case, this is warranted by taking the
central extension of $G$ or of its Lie algebra $\mathfrak{g}$, providing
intersections with a rich class of coadjoint orbits inside. However, this
seems to be a very specific situation, in general it happens that the momentum
maps have no non trivial intersection, as it is the case in finite dimension.

Handling this problem in a general fashion lead us to propose a wider scheme
for PL T-duality by considering T-dual equivalence classes constructed
alternatively on hamiltonian $G$, $K$ or $B$-spaces. As we shall show below,
the main facts underlying the standard PL T-duality remain the same: the
canonical transformation between systems on the factors $K$ and $B$ arises
from the symmetries involving their \emph{Poisson-Lie structure}. In this way,
one is able to built up PL T-dual equivalence classes attached to coadjoint
orbits in $\mathfrak{g}^{\ast}$, $\mathfrak{k}^{\ast}$or $\mathfrak{b}^{\ast}%
$. In addition, this wider framework allows to make contact with the
\emph{AKS} theory for integrable systems.

So, let us consider the Lie group $G$ and its Iwasawa decomposition $G=KB$,
where $K$ is the compact factor and $B$ is the solvable one. The abstract
framework we use here also includes the Lie algebras $\mathfrak{g}%
,\mathfrak{k},\mathfrak{b}$, which correspond to the Lie groups $G,K,B$ so
that $\mathfrak{g}=\mathfrak{k}\oplus\mathfrak{b}$, and $\mathfrak{g}$ is
equipped with a non degenerate symmetric bilinear form $\left(  ,\right)
_{\mathfrak{g}}$ turning $\mathfrak{k}$ and $\mathfrak{b}$ into isotropic
subspaces. This allows the identification $\mathfrak{b}^{\ast}\simeq
\mathfrak{k}$ and $\mathfrak{k}^{\ast}\simeq\mathfrak{b}$. The projectors are
denoted by $\Pi_{K}:G\rightarrow K$, $\Pi_{B}:G\rightarrow B$; moreover, the
symbols $\Pi_{\mathfrak{k}}:\mathfrak{g}\rightarrow\mathfrak{k},\Pi
_{\mathfrak{b}}:\mathfrak{g}\rightarrow\mathfrak{b}$ are meant to indicate the
projections into the summands of the Lie algebra decomposition induced by the factorization.

Let us describe a PL T-duality scheme based on the the action of one of the
factors, $B$ in this case, instead of the action of $G$. It will involve the
Poisson manifold $\left(  \mathfrak{b}^{\ast},\left\{  ,\right\}
_{\mathfrak{b}^{\ast}}\right)  $, where $\left\{  ,\right\}  _{\mathfrak{b}%
^{\ast}}$ is the Kirillov-Kostant bracket, and the symplectic manifolds
$\left(  T^{\ast}K,\omega_{K}\right)  $ and $\left(  T^{\ast}B,\omega
_{B}\right)  $, with $\omega_{K}$,$\omega_{B}$ standing for the canonical
symplectic forms on each phase spaces, respectively. All the cotangent bundles
are regarded in body coordinates, so they are trivialized by left translations.

The phase space $T^{\ast}B\cong B\times\mathfrak{b}^{\ast}$ turns in a
hamiltonian $B$-space by the action $\tau:$ $B\times\left(  B\times
\mathfrak{b}^{\ast}\right)  \longrightarrow B\times\mathfrak{b}^{\ast}$
obtained as the lift of the action of $B$ on itself by left translations%
\[
\tau\left(  \tilde{g},(\tilde{h},\tilde{\eta})\right)  =(\tilde{g}\tilde
{h},\tilde{\eta})
\]
for $\tilde{g},\tilde{h}\in B$, $\tilde{\eta}\in\mathfrak{b}^{\ast}$, with
$\mathrm{Ad}$-equivariant momentum map $\lambda:B\times\mathfrak{b}^{\ast
}\longrightarrow\mathfrak{b}^{\ast}$
\[
\lambda(\tilde{h},\tilde{\eta})=\mathrm{Ad}_{\tilde{h}^{-1}}^{\ast}\tilde
{\eta}%
\]

On the other side, $T^{\ast}K\cong K\times\mathfrak{k}^{\ast}$, becomes in a
hamiltonian $B$-space by virtue of the \emph{Poisson-Lie structure} of $K$
inherited from the Iwasawa decomposition $G=KB$. In fact, we introduce the
action $\mathsf{dr}:B\times\left(  K\times\mathfrak{k}^{\ast}\right)
\longrightarrow K\times\mathfrak{k}^{\ast}$%
\begin{equation}
\mathsf{dr}\left(  \tilde{b},\left(  g,\eta\right)  \right)  =\left(
g^{\tilde{b}},\mathrm{Ad}_{\tilde{b}^{g}}\eta\right)  \label{dress-action}%
\end{equation}
for $\tilde{b}\in B$, $g\in K$, $\eta\in\mathfrak{k}^{\ast}$ which is obtained
by lifting the \emph{dressing} \emph{action }of $B$ on $K$ to the cotangent
bundle. This action, introduced in $\cite{STS},\cite{Lu-We}$, works as
follows: by writing each element $l\in G$ as $l=g\tilde{h}$, with $g\in K$ and
$\tilde{h}\in B^{\ast}$, the product $\tilde{h}g$ in $G$ can be expressed as
$\tilde{h}g=g^{\tilde{h}}\tilde{h}^{g}$, with $g^{\tilde{h}}\in K$ and
$\tilde{h}^{g}\in B$. The \emph{dressing action} of $B$ on $K$ is defined as
$\mathsf{Dr}:B\times K\longrightarrow K$, such that $\mathsf{Dr}\left(
\tilde{h},g\right)  :=\Pi_{K}\tilde{h}g=g^{\tilde{h}}$. Infinitesimally,
$\xi\in\mathfrak{b}$ is mapped onto the tangent vector
\[
\left(  \xi_{K\times\mathfrak{k}^{\ast}}\right)  _{\left(  g,\eta\right)
}=\left(  g^{\xi},\left[  Ad_{g}^{\ast}\xi,\eta\right]  \right)
\]
at $\left(  g,\eta\right)  \in K\times\mathfrak{k}^{\ast}$; here are assumed
the identifications explained above, so that $Ad_{g}^{\ast}:\mathfrak{b}%
\rightarrow\mathfrak{b}$ and the bracket is the Lie bracket in $\mathfrak{b}$.
The momentum map $\phi:K\times\mathfrak{k}^{\ast}\longrightarrow
\mathfrak{b}^{\ast}$ is
\[
\phi(g,\eta)=g\left(  g^{-1}\right)  ^{\eta}%
\]
having in mind that the right hand side belongs to $\mathfrak{k}%
\simeq\mathfrak{b}^{\ast}$. In order to avoid confusion, these identifications
will be explicitly shown in the specific case addressed in the following sections.

The momentum maps $\phi$ and $\lambda$ turn $\left(  K\times\mathfrak{k}%
^{\ast},\omega_{K}\right)  $ and $\left(  B\times\mathfrak{b}^{\ast}%
,\omega_{B}\right)  $ in symplectic realizations of the Poisson manifold
$\left(  \mathfrak{b}^{\ast},\left\{  ,\right\}  _{\mathfrak{b}^{\ast}%
}\right)  $, as depicted in the diagram
\begin{equation}
\begin{diagram} \node{K\times\mathfrak{k}^*}\arrow{se,b}{\phi}\node[2]{B\times\mathfrak{b}^*}\arrow{sw,r}{\lambda}\\ \node[2]{\mathfrak{b}^*} \end{diagram} \label{T-duality scheme 0}%
\end{equation}
which is the basic geometric scheme underlying PL T-duality. Seeking for
compatible dynamics drives to the realm of \emph{collective hamiltonian
systems }$\cite{Guill-Sten}$, meaning that a hamiltonian function
$\mathsf{h}\in C^{\infty}\left(  \mathfrak{b}^{\ast}\right)  $ is the
masterpiece governing both the PL T-dual systems on $K\times\mathfrak{k}%
^{\ast}$ and $B\times\mathfrak{b}^{\ast}$. In fact, the corresponding pull
backs by the momentum maps $\phi$ and $\lambda$, namely $\mathsf{h}\circ
\phi\in C^{\infty}\left(  K\times\mathfrak{k}^{\ast}\right)  $ and
$\mathsf{h}\circ\lambda\in C^{\infty}\left(  B\times\mathfrak{b}^{\ast
}\right)  $, produce the desired compatible dynamics.

These systems are said to be in \emph{collective hamiltonian form} and to
understand its geometric meaning we work on a generic hamiltonian $B$-space
$(M,\omega)$, with an $\mathrm{Ad}$-equivariant momentum map $J:M\rightarrow
\mathfrak{b}^{\ast}$ associated with the symplectic action $\varphi:B\times
M\longrightarrow M$ of the Lie group $B$, and taking the \emph{collective
hamiltonian} $H=\mathsf{h}\circ J$. In terms of the \emph{orbit map} through
$m\in M$, $\varphi_{m}:B\longrightarrow M/\varphi_{m}\left(  b\right)
:=\varphi\left(  b,m\right)  $, the infinitesimal generators can be written as
$X_{M}\left(  m\right)  =\left(  \varphi_{m}\right)  _{\ast}X$, for
$X\in\mathfrak{b}$ and $X_{M}\in\mathfrak{X}\left(  M\right)  $. Hence,
introducing the \emph{Legendre transformation }of\emph{ }$\mathsf{h}$, namely
the linear map $\mathcal{L}_{\mathsf{h}}:\mathfrak{b}^{\ast}\rightarrow
\mathfrak{b}$ defined as $\mathbf{\langle}\xi,\mathcal{L}_{\mathsf{h}}%
(\eta)\mathbf{\rangle}_{\mathfrak{g}}=\mathbf{\langle}\left.  d\mathsf{h}%
\right\vert _{\eta},\xi\mathbf{\rangle}$, for any $\xi\in\mathfrak{b}^{\ast}$,
we may write the hamiltonian vector field of $H$ as\textit{ }
\[
\left.  V_{H}\right\vert _{m}=\left(  \varphi_{m}\right)  _{\ast}\left[
\mathcal{L}_{\mathsf{h}}\circ J\right]  (m)
\]
and its image by $J$ is tangent to the coadjoint orbit through\textit{
}$J\left(  m\right)  $
\[
\left.  J_{\ast}\right\vert _{m}V_{H}=-\left(  \mathrm{ad}_{\mathcal{L}%
_{\mathsf{h}}\left(  J(m)\right)  }^{B}\right)  ^{\ast}J\left(  m\right)
\]
In other words, the hamiltonian vector field $V_{H}$ is mapped on the tangent
space of a coadjoint orbits in $\mathfrak{b}^{\ast}$. If $m(t)$ denotes the
trajectory of the hamiltonian system through $m(0)=m$, $\dot{m}(t)=\left.
V_{H}\right\vert _{m(t)}$, the images $\gamma(t)=J\left(  m(t)\right)  $ lies
completely on the coadjoint orbit through $J\left(  m\right)  $, where the
equation of motion is
\begin{equation}
\dot{\gamma}(t)=-\left(  \mathrm{ad}_{\mathcal{L}_{\mathsf{h}}\left(
\gamma(t)\right)  }^{B}\right)  ^{\ast}\gamma(t) \label{SolHamSistDual}%
\end{equation}
that corresponds to a hamiltonian system on the coadjoint orbits on
$\mathfrak{b}^{\ast}$, with hamiltonian function $\mathsf{h}$.

\begin{description}
\item[Proposition:] \label{PropFundamentalAKSTduality}\textit{Let }%
$\gamma:\mathbb{R}\rightarrow\mathfrak{b}^{\ast}$\textit{\ be the solution
curve of Eq. }$\left(  \ref{SolHamSistDual}\right)  $\textit{ with initial
condition }$\gamma\left(  0\right)  =J_{M}\left(  m\right)  $\textit{, and
select a curve }$b\left(  t\right)  $\textit{\ in }$B$\textit{\ such that }
\begin{equation}
\gamma(t)=\left(  \mathrm{Ad}_{b^{-1}\left(  t\right)  }^{B}\right)  ^{\ast
}J_{M}\left(  m\right)  \label{DefGroupCurve}%
\end{equation}
\textit{Then, among these curves there exists a solution of the differential
equation on }$B$%
\begin{equation}%
\begin{array}
[c]{ccc}%
\dot{b}(t)b^{-1}(t)=\mathcal{L}_{\mathsf{h}}\left(  \gamma(t)\right)  & , &
b(0)=n_{0}\in B_{J_{M}\left(  m\right)  }%
\end{array}
\label{DiffEqonB}%
\end{equation}
\textit{where }$B_{J_{M}\left(  m\right)  }$\textit{ is the stabilizer group
of the point }$J_{M}\left(  m\right)  $\textit{ under the coadjoint action of
}$B$\textit{ on }$\mathfrak{b}^{\ast}$.
\end{description}

\textbf{Proof}: Let us suppose that $b:\mathbb{R}\rightarrow B$ satisfies Eq.
$\left(  \ref{SolHamSistDual}\right)  $ through Eq. \eqref{DefGroupCurve}, and
take $n:\mathbb{R}\rightarrow B_{J_{M}\left(  m\right)  }$ such that $b\left(
t\right)  n\left(  t\right)  $ solves the differential equation $\left(
\ref{DiffEqonB}\right)  $. Then
\[
\frac{d\left(  b\left(  t\right)  n\left(  t\right)  \right)  }{dt}\left(
b\left(  t\right)  n\left(  t\right)  \right)  ^{-1}=\mathcal{L}_{\mathsf{h}%
}\left(  \gamma\left(  t\right)  \right)
\]
or equivalently
\[
\dot{n}\left(  t\right)  n^{-1}\left(  t\right)  =\mathrm{Ad}_{b^{-1}\left(
t\right)  }^{B}\mathcal{L}_{\mathsf{h}}\left(  \gamma\left(  t\right)
\right)  -b^{-1}\left(  t\right)  \dot{b}\left(  t\right)
\]
We have to verify that the right hand side of this expression belongs to
$\mathfrak{b}_{J_{M}\left(  m\right)  }$, the Lie algebra of the stabilizer
subgroup $B_{J_{M}\left(  m\right)  }$. Taking into account that $b$ satisfies
Eq. $\left(  \ref{DefGroupCurve}\right)  $, we have
\[
\dot{b}\left(  t\right)  b^{-1}\left(  t\right)  =\mathcal{L}_{\mathsf{h}%
}\left(  \gamma\left(  t\right)  \right)  +M\left(  t\right)
\]
for some curve $M:\mathbb{R}\rightarrow\mathfrak{b}_{\gamma\left(  t\right)
}$. Furthermore, we have that $X\in\mathfrak{b}_{\gamma\left(  t\right)  }$
iff $\left(  \mathrm{ad}_{X}^{B}\right)  ^{\ast}\gamma\left(  t\right)  =0$,
and this means
\[
0=\left(  \mathrm{ad}_{X}^{B}\right)  ^{\ast}\left(  \mathrm{Ad}%
_{b^{-1}\left(  t\right)  }^{B}\right)  ^{\ast}J_{M}\left(  m\right)  =\left(
\mathrm{Ad}_{b^{-1}\left(  t\right)  }^{B}\right)  ^{\ast}\left(
\mathrm{ad}_{\mathrm{Ad}_{b^{-1}\left(  t\right)  }X}^{B}\right)  ^{\ast}%
J_{M}\left(  m\right)  .
\]
Then $X\in\mathfrak{b}_{\gamma\left(  t\right)  }$ iff $\mathrm{Ad}%
_{b^{-1}\left(  t\right)  }^{B}X\in\mathfrak{b}_{J_{M}\left(  m\right)  }$.
Therefore, $M\left(  t\right)  =\mathrm{Ad}_{b\left(  t\right)  }^{B}N\left(
t\right)  $ for some curve $N:\mathbb{R}\rightarrow\mathfrak{b}_{J_{M}\left(
m\right)  }$, and finally
\[
\dot{n}\left(  t\right)  n^{-1}\left(  t\right)  =N\left(  t\right)
\in\mathfrak{b}_{J_{M}\left(  m\right)  }%
\]
as we want to show.$\blacksquare$

Hence, $m(t)=\varphi\left(  b(t),m\right)  $ is the solution to the original
hamiltonian system. Moreover, if $\mathfrak{b}$ is supplied with an invariant
non degenerate bilinear form $\left(  \,,\right)  :\mathfrak{b}\times
\mathfrak{b}\longrightarrow\mathbb{K}$, and denoting $\tilde{\gamma
}:\mathbb{R}\rightarrow\mathfrak{b}$ the image of $\gamma:\mathbb{R}%
\rightarrow\mathfrak{b}^{\ast}$ through the induced isomorphism $\mathfrak{b}%
^{\ast}\rightarrow\mathfrak{b}$, the equation of motion turns into the Lax
form
\begin{equation}
\dfrac{d\tilde{\gamma}(t)}{dt}=[\tilde{\gamma}(t),\mathcal{L}_{\mathsf{h}%
}\left(  \gamma(t)\right)  ] \label{163}%
\end{equation}

\subsection{Relation with AKS method}

The success of the method described above relies on the integrability of the
Eq. $\left(  \ref{SolHamSistDual}\right)  $. The AKS theory $\cite{AKS}$ gives
a family of integrable hamiltonians associated to $\mathrm{Ad}^{G\ast}%
$-invariant functions on $\mathfrak{g}^{\ast}$. First, we have the
identification $\mathfrak{k}^{\circ}\simeq\mathfrak{b}^{\ast}$ by the map
$\eta\in\mathfrak{b}^{\ast}\mapsto\eta\circ\Pi_{\mathfrak{b}}$. It allows us
to define a $B$-action via
\[
\tau_{b}^{B}\left(  \mu\right)  :=\left(  \left(  \mathrm{Ad}_{b}^{G}\right)
^{\ast}\mu\right)  \circ\Pi_{\mathfrak{b}}\qquad\forall b\in B,\mu
\in\mathfrak{k}^{\circ}%
\]
The orbit $\mathcal{O}_{\mu}^{B}\subset\mathfrak{k}^{\circ}$ for this action
through $\mu$ is a symplectic manifold; in fact, for $\eta\in\mathcal{O}_{\mu
}^{B}$ we have that
\[
T_{\eta}\left(  \mathcal{O}_{\mu}^{B}\right)  =\left\{  \left(  \mathrm{ad}%
_{X}^{G}\right)  ^{\ast}\eta\circ\Pi_{\mathfrak{b}}:X\in\mathfrak{b}\right\}
\]
and the symplectic structure is given by
\[
\left\langle \omega,\left(  \mathrm{ad}_{X_{1}}^{G}\right)  ^{\ast}\eta
\circ\Pi_{\mathfrak{b}}\otimes\left(  \mathrm{ad}_{X_{2}}^{G}\right)  ^{\ast
}\eta\circ\Pi_{\mathfrak{b}}\right\rangle _{\eta}=\left\langle \eta,\left[
X_{1},X_{2}\right]  \right\rangle
\]
This structure will be used in proving the following result.

\begin{description}
\item[Theorem] \textit{Let }$f\in C^{\infty}\left(  \mathfrak{g}^{\ast
}\right)  $\textit{ be an }$\mathrm{Ad}^{\ast}$-\textit{invariant function,
and let the restriction }$\mathsf{h}:=\left.  f\right\vert \mathcal{O}%
_{J_{M}\left(  m\right)  }^{B}$ \textit{be the hamiltonian function for the
system defined on }$\mathcal{O}_{J_{M}\left(  m\right)  }^{B}\subset
\mathfrak{k}^{\circ}\simeq\mathfrak{b}^{\ast}$\textit{. Hence, the solution of
this system with initial condition }$\eta\left(  0\right)  =J_{M}\left(
m\right)  $\textit{ is }
\[
\eta\left(  t\right)  =\left(  \mathrm{Ad}_{k\left(  t\right)  }^{G}\right)
^{\ast}J_{M}\left(  m\right)
\]
\textit{where }$k:\mathbb{R}\rightarrow K$\textit{ is the }$K$-\textit{factor
in the decomposition of the element }$g\left(  t\right)  =\exp\left(
t\mathcal{L}_{f}\left(  J_{M}\left(  m\right)  \right)  \right)  $.
\end{description}

\textbf{Proof:} Let $\eta$ be an arbitrary element in the orbit $\mathcal{O}%
_{J_{M}\left(  m\right)  }^{B}$ defined above. In this case, using the
associated Legendre transformation $\mathcal{L}_{f}:\mathfrak{g}^{\ast
}\longrightarrow\mathfrak{g}$ that allows to identify $T_{\eta}^{\ast}\left(
\mathfrak{g}^{\ast}\right)  \simeq\mathfrak{g}$, we have
\[
\left\langle df,\left(  \mathrm{ad}_{X}^{G}\eta\right)  ^{\ast}\circ
\Pi_{\mathfrak{b}}\right\rangle _{\eta}=\left\langle \left(  \mathrm{ad}%
_{X}^{G}\eta\right)  ^{\ast}\circ\Pi_{\mathfrak{b}},\mathcal{L}_{f}\left(
\eta\right)  \right\rangle =\left\langle \eta,\left[  X,\Pi_{\mathfrak{b}%
}\mathcal{L}_{f}\left(  \eta\right)  \right]  \right\rangle
\]
so that the hamiltonian vector field associated to $\mathsf{h}$ is given by
\[
\left.  V_{\mathsf{h}}\right\vert _{\eta}=-\left(  \mathrm{ad}_{\Pi
_{\mathfrak{b}}\mathcal{L}_{f}\left(  \eta\right)  }^{G}\eta\right)  ^{\ast
}\circ\Pi_{\mathfrak{b}}.
\]
Because of the $\mathrm{Ad}^{\ast}$-invariance of $f$, we have that $\left(
\mathrm{ad}_{\mathcal{L}_{f}\left(  \eta\right)  }^{G}\right)  ^{\ast}\eta=0$,
and we can rewrite it as
\begin{equation}
\left.  V_{\mathsf{h}}\right\vert _{\eta}=\left(  \mathrm{ad}_{\Pi
_{\mathfrak{k}}\mathcal{L}_{f}\left(  \eta\right)  }^{G}\right)  ^{\ast}%
\eta\label{HamVectorFieldProof}%
\end{equation}
by taking into account that $\left(  \mathrm{Ad}_{k}^{G}\right)  ^{\ast
}\mathfrak{k}^{\circ}\subset\mathfrak{k}^{\circ}$ for all $k\in K$.

On the other side, the curve in $\mathfrak{k}^{\circ}$ defined through
\[
\eta\left(  t\right)  =\left(  \mathrm{Ad}_{k\left(  t\right)  }^{G}\right)
^{\ast}J_{M}\left(  m\right)
\]
has tangent vector field given by%
\[
\dot{\eta}\left(  t\right)  =\left.  \vec{\frac{d}{dt}}\right\vert _{t}\left[
\left(  \mathrm{Ad}_{k\left(  t\right)  }^{G}\right)  ^{\ast}J_{M}\left(
m\right)  \right]  =\left(  \mathrm{ad}_{k^{-1}\left(  t\right)  \dot
{k}\left(  t\right)  }^{G}\right)  ^{\ast}\eta\left(  t\right)  .
\]
Now, considering the integral curve $g\left(  t\right)  $ of the right
invariant vector field $\mathcal{L}_{f}\left(  J_{M}\left(  m\right)  \right)
g\left(  t\right)  $, written in terms of the decomposition curves $g\left(
t\right)  =k\left(  t\right)  b\left(  t\right)  $ we obtain%
\[
\dot{g}\left(  t\right)  g^{-1}\left(  t\right)  =\left.  \frac{d\left(
kb\right)  }{dt}\left(  kb\right)  ^{-1}\right\vert _{t}=\mathrm{Ad}_{k\left(
t\right)  }^{G}\left(  k^{-1}\left(  t\right)  \dot{k}\left(  t\right)
+\dot{b}\left(  t\right)  b^{-1}\left(  t\right)  \right)
\]
meaning that $\mathrm{Ad}_{k^{-1}\left(  t\right)  }^{G}\mathcal{L}_{f}\left(
J_{M}\left(  m\right)  \right)  =k^{-1}\left(  t\right)  \dot{k}\left(
t\right)  +\dot{b}\left(  t\right)  b^{-1}\left(  t\right)  $, and therefore
\[
\Pi_{\mathfrak{k}}\left(  \mathrm{Ad}_{k^{-1}\left(  t\right)  }%
^{G}\mathcal{L}_{f}\left(  J_{M}\left(  m\right)  \right)  \right)
=k^{-1}\left(  t\right)  \dot{k}\left(  t\right)
\]
By using $\mathrm{Ad}^{\ast}$-invariance for $f$ again, we have that
\[
\mathrm{Ad}_{k^{-1}\left(  t\right)  }^{G}\mathcal{L}_{f}\left(  J_{M}\left(
m\right)  \right)  =\mathcal{L}_{f}\left(  \left[  \mathrm{Ad}_{k\left(
t\right)  }^{G}\right]  ^{\ast}J_{M}\left(  m\right)  \right)  =\mathcal{L}%
_{f}\left(  \eta\left(  t\right)  \right)
\]
implying that $k^{-1}\left(  t\right)  \dot{k}\left(  t\right)  =\Pi
_{\mathfrak{k}}\left(  \mathcal{L}_{f}\left(  \eta\left(  t\right)  \right)
\right)  $. Comparing with Eq. $\left(  \ref{HamVectorFieldProof}\right)  $ we
can conclude that $\eta\left(  t\right)  $ has $V_{\mathsf{h}}$ as tangent
vector field. $\blacksquare$

The $\mathrm{Ad}^{\ast}$-invariance implies the identity $\left(
\mathrm{ad}_{\mathcal{L}_{f}\left(  \eta\right)  }^{G}\right)  ^{\ast}\eta=0$,
for all $\eta\in\mathfrak{g}^{\ast}$, and so
\[
\left[  \mathrm{Ad}_{\exp t\mathcal{L}_{f}\left(  J_{M}\left(  m\right)
\right)  }^{G}\right]  ^{\ast}J_{M}\left(  m\right)  =J_{M}\left(  m\right)
\]
meaning that
\[
\left[  \mathrm{Ad}_{k\left(  t\right)  }^{G}\right]  ^{\ast}J_{M}\left(
m\right)  =\left[  \mathrm{Ad}_{b^{-1}\left(  t\right)  }^{G}\right]  ^{\ast
}J_{M}\left(  m\right)
\]
and assuming $J_{M}\left(  m\right)  \in\mathfrak{k}^{\circ}$ it is clear that
we can take $b:\mathbb{R}\rightarrow B$ (the $B$-factor of $\exp
t\mathcal{L}_{f}\left(  J_{M}\left(  m\right)  \right)  $) as the solution
curve in Eq. $\left(  \ref{SolHamSistDual}\right)  $. In such case it is
necessary to find the differential equation for the $B_{J_{M}\left(  m\right)
}$-factor $n$ (cf. proof of the proposition \ref{PropFundamentalAKSTduality}
above). But as was previously shown
\[
\mathcal{L}_{f}\left(  \eta\left(  t\right)  \right)  =k^{-1}\left(  t\right)
\dot{k}\left(  t\right)  +\dot{b}\left(  t\right)  b^{-1}\left(  t\right)
,\qquad\eta\left(  t\right)  =\mathrm{Ad}_{k^{-1}\left(  t\right)  }^{G}%
J_{M}\left(  m\right)  ,
\]
so $\dot{b}\left(  t\right)  b^{-1}\left(  t\right)  =\Pi_{\mathfrak{b}%
}\mathcal{L}_{f}\left(  \eta\left(  t\right)  \right)  $.

On the other side, for all $\xi,\eta\in\mathfrak{k}^{\circ}$ we have that%
\[
\left\langle \mathcal{L}_{f}\left(  \eta\right)  ,\xi\right\rangle
=\left\langle \Pi_{\mathfrak{b}}\mathcal{L}_{f}\left(  \eta\right)
,\xi\right\rangle
\]%
\[
\left\langle \xi,\mathcal{L}_{f}\left(  \eta\right)  \right\rangle =\left.
\frac{d}{dt}f\left(  \eta+t\xi\right)  \right\vert _{t=0}=\left.  \frac{d}%
{dt}\mathsf{h}\left(  \eta+t\xi\right)  \right\vert _{t=0}=\left\langle
\xi,\mathcal{L}_{\mathsf{h}}\left(  \eta\right)  \right\rangle
\]
meaning that $\mathcal{L}_{\mathsf{h}}\left(  \eta\right)  =\Pi_{\mathfrak{b}%
}\mathcal{L}_{f}\left(  \eta\right)  $ and then $\dot{b}\left(  t\right)
b^{-1}\left(  t\right)  =\mathcal{L}_{\mathsf{h}}\left(  \eta\right)  $;
therefore
\[
\dot{n}\left(  t\right)  n^{-1}\left(  t\right)  =0
\]
and the $B_{J_{M}\left(  m\right)  }$-factor is constant.

\subsection{Summary}

The setting consist of a factorizable Lie group $G=KB$ and a hamiltonian
$B$-space $M$. The collective motion associated to the restriction to
$\mathfrak{k}^{\circ}\simeq\mathfrak{b}^{\ast}$ of an $\left(  \mathrm{Ad}%
^{G}\right)  ^{\ast}$-invariant function $f$ gives rise to a collective
hamiltonian system on $M$, which can be thus solved algebraically as follows:

\begin{enumerate}
\item Factorize the straight curve $t\mapsto\exp{t}\mathcal{L}_{f}\left(
J_{M}\left(  m\right)  \right)  =k\left(  t\right)  b\left(  t\right)  $.

\item The solution curve on $M$ for the hamiltonian system defined by
$H:=\left(  \left.  f\right\vert \mathfrak{k}^{\circ}\right)  \circ J_{M}$ is
given by
\[
t\mapsto\varphi\left(  b\left(  t\right)  n_{0},m\right)
\]
for some element $n_{0}\in B_{J_{M}\left(  m\right)  }$.
\end{enumerate}

\section{Iwasawa decomposition of $SL(2,\mathbb{C})$ and coadjoint orbits}

We now specialize the above abstract structure to $G=SL(2,\mathbb{C})$ and its
Iwasawa decomposition $SL(2,\mathbb{C})\cong SU\left(  2\right)  \times B$,
where $B$ is the solvable group of $2\times2$ complex upper triangular
matrices, with real positive diagonals and determinant equal to $1$. Let us
address the construction of an explicit example of T-dual systems in this framework.

In order to start with, we consider the maximal Abelian subalgebra
$\mathfrak{h}=\mathbb{C}\langle\sigma_{3}\rangle$ of the Lie algebra
$\mathfrak{sl}_{2}(\mathbb{C})$, with the root system $\Delta:=\{-\alpha
,\alpha\}$, where $\alpha\in\mathfrak{h}^{\ast}$ is given by $\alpha
(\sigma_{3})=2$. The associated decomposition is $\mathfrak{sl}_{2}%
(\mathbb{C})=\mathfrak{h}\oplus\mathfrak{g}_{\alpha}\oplus\mathfrak{g}%
_{-\alpha}$, with
\[
\mathfrak{g}_{\pm\alpha}:=\mathbb{C}\langle\frac{1}{2}(\sigma_{1}\pm
i\sigma_{2})\rangle.
\]
For the positive root set $\Delta_{+}=\{\alpha\}$ we define $\mathfrak{n}%
:=\bigoplus_{\beta\in\Delta_{+}}\mathfrak{g}_{\beta}=g_{\alpha}$. Then we may
find a decomposition as expected for $\mathfrak{sl}_{2}(\mathbb{C}%
)^{\mathbb{R}}$ by taking $\mathfrak{k}=\mathfrak{su}_{2}$ and $\mathfrak{b}%
:=\mathfrak{a}\oplus\mathfrak{n}^{\mathbb{R}}$, where $\mathfrak{a}%
:=\mathbb{R}\langle\sigma_{3}\rangle=i\mathfrak{t}$, being $\mathfrak{t}%
:=\mathfrak{h}\cap\mathfrak{su}_{2}$ a real form for $\mathfrak{h}$,%
\[
\mathfrak{sl}_{2}(\mathbb{C})^{\mathbb{R}}=\mathfrak{su}_{2}\oplus\mathfrak{b}%
\]
With this election for $\mathfrak{h}$, $\mathfrak{b}$ is the subalgebra of
upper triangular matrices with real diagonal and null trace, and
$\mathfrak{k}$ is the real subalgebra of $\mathfrak{sl}_{2}(\mathbb{C})$ of
antihermitean matrices.

Alternatively, one would may choose for instance $\mathfrak{h}^{\prime
}:=\mathbb{C}\langle\sigma_{1}\rangle$, changing the roots $\alpha^{\prime}$
and the spaces $\mathfrak{g}_{\alpha^{\prime}}$, so that $\mathfrak{b}%
^{\prime}$ is no longer composed of upper triangular matrices. However, by
change of basis (the one which diagonalize $\sigma_{1}$) will turn
$\mathfrak{b}^{\prime}$ into triangular matrices again. The compact real form
is obtained as usual, defining
\[
\mathfrak{u}_{n}:=\sum_{\alpha\in\Delta}\mathbb{R}(iH_{\alpha})+\sum
_{\alpha\in\Delta}\mathbb{R}(X_{\alpha}-X_{-\alpha})+\sum_{\alpha\in\Delta
}\mathbb{R}i(X_{\alpha}+X_{-\alpha})
\]
once $\mathfrak{h}$ is fixed.

The Killing form for $\mathfrak{sl}_{2}(\mathbb{C})$ is $\kappa
(X,Y):=\mathsf{tr}\,{(ad}\left(  {X}\right)  {{ad}}\left(  {Y}\right)
{)}=4\mathsf{tr}\,{(XY)}$, the restrictions to $\mathfrak{su}_{2}$,
$\mathfrak{a}$, and $\mathfrak{n}$ are negative defined, positive defined, and
$0$, respectively. We consider the non degenerate symmetric bilinear form on
$\mathfrak{sl}_{2}(\mathbb{C})$
\begin{equation}
(X,Y)_{\mathfrak{sl}_{2}}=-\frac{1}{4}\mathrm{\operatorname{Im}}\,\kappa(X,Y)
\label{prod-escal 0}%
\end{equation}
which turns $\mathfrak{b}$ and $\mathfrak{k}$ into isotropic subspaces. Also,
we take the basis
\[%
\begin{array}
[c]{ccccc}%
X_{1}=\left[
\begin{array}
[c]{cc}%
0 & i\\
i & 0
\end{array}
\right]  &  & X_{2}=\left[
\begin{array}
[c]{cc}%
0 & 1\\
-1 & 0
\end{array}
\right]  &  & X_{3}=\left[
\begin{array}
[c]{cc}%
i & 0\\
0 & -i
\end{array}
\right]
\end{array}
\]
for $\mathfrak{su}_{2}$, and%
\[%
\begin{array}
[c]{ccccc}%
E=\left[
\begin{array}
[c]{cc}%
0 & 1\\
0 & 0
\end{array}
\right]  ~, &  & iE=\left[
\begin{array}
[c]{cc}%
0 & i\\
0 & 0
\end{array}
\right]  ~, &  & H=\left[
\begin{array}
[c]{cc}%
1 & 0\\
0 & -1
\end{array}
\right]
\end{array}
\]
in $\mathfrak{b}$. Then, the crossed product are
\begin{equation}%
\begin{array}
[c]{ccccc}%
(X_{1},E)_{\mathfrak{sl}_{2}}=-1 &  & (X_{2},E)_{\mathfrak{sl}_{2}}=0 &  &
(X_{3},E)_{\mathfrak{sl}_{2}}=0\\
(X_{1},iE)_{\mathfrak{sl}_{2}}=0 &  & (X_{2},iE)_{\mathfrak{sl}_{2}}=1 &  &
(X_{3},iE)_{\mathfrak{sl}_{2}}=0\\
(X_{1},H)_{\mathfrak{sl}_{2}}=0 &  & (X_{1},H)_{\mathfrak{sl}_{2}}=0 &  &
(X_{3},H)_{\mathfrak{sl}_{2}}=-2
\end{array}
\label{prod-escal}%
\end{equation}
allowing for the identification $\psi:\mathfrak{su}_{2}\rightarrow
\mathfrak{b}^{\ast}$ given by
\begin{equation}
\psi(X_{1})=-\mathbf{e},\qquad\psi(X_{2})=\mathbf{\tilde{e}},\qquad\psi
(X_{3})=-2\mathbf{h} \label{psi}%
\end{equation}
where $\left\{  \mathbf{e},\mathbf{\tilde{e}},\mathbf{h}\right\}
\subset\mathfrak{b}^{\ast}$ is the dual basis to $\left\{  E,iE,H\right\}
\subset\mathfrak{b}$.

This map allows to carry the Poisson structure of $\mathfrak{b}^{\ast}$ to
$\mathfrak{su}_{2}$. In terms of the dual basis $\{\mathbf{x}_{k}%
\}\subset\mathfrak{su}_{2}^{\ast}$, $\left\langle \mathbf{x}_{k}%
,X_{j}\right\rangle =\delta_{kj}$, so for $f\in C^{\infty}\left(
\mathfrak{su}_{2}\right)  $ we have that $df(X_{k})=\dfrac{\partial
f}{\partial\mathbf{x}_{k}}$ and the Poisson bracket reads as
\[
\{f,g\}=\left(  \frac{\partial f}{\partial\mathbf{x}_{1}}\frac{\partial
g}{\partial\mathbf{x}_{3}}-\frac{\partial f}{\partial\mathbf{x}_{3}}%
\frac{\partial g}{\partial\mathbf{x}_{1}}\right)  \mathbf{x}_{1}+\left(
\frac{\partial f}{\partial\mathbf{x}_{2}}\frac{\partial g}{\partial
\mathbf{x}_{3}}-\frac{\partial f}{\partial\mathbf{x}_{3}}\frac{\partial
g}{\partial\mathbf{x}_{2}}\right)  \mathbf{x}_{2}.
\]
The hamiltonian vector fields are then%
\[
V_{g}=\frac{\partial g}{\partial\mathbf{x}_{3}}\left(  \mathbf{x}_{1}%
\frac{\partial}{\partial\mathbf{x}_{1}}+\mathbf{x}_{2}\frac{\partial}%
{\partial\mathbf{x}_{2}}\right)  -\left(  \mathbf{x}_{1}\frac{\partial
g}{\partial\mathbf{x}_{1}}+\mathbf{x}_{2}\frac{\partial g}{\partial
\mathbf{x}_{2}}\right)  \frac{\partial}{\partial\mathbf{x}_{3}}%
\]
With the identification $X_{k}=\dfrac{\partial}{\partial\mathbf{x}_{k}}$, we
get
\[
X_{\mathbf{x}_{1}}=\mathbf{x}_{1}X_{3},\qquad X_{\mathbf{x}_{2}}%
=\mathbf{x}_{2}X_{3},\qquad X_{\mathbf{x}_{3}}=-\mathbf{x}_{1}X_{1}%
-\mathbf{x}_{2}X_{2},
\]
from where it can be determined the \emph{symplectic leaves}, which are
divided in two uniparametric families, namely,

\begin{description}
\item[-] \textit{Symplectic leaves of} \textit{dimension} $0$\textit{: each
leaf is a point }$\alpha X_{3},\alpha\in\mathbb{R}$\textit{,} \textit{on the
vertical axis of }$\mathfrak{su}_{2}\simeq\mathbb{R}^{3}$,

\item[-] \textit{Symplectic leaves of dimension} $2$\textit{: each leaf is a
vertical semiplane }%
\begin{equation}
\mathcal{O}_{\theta}=\left\{  \left(  xX_{\theta}+zX_{3}\right)
\in\mathfrak{su}_{2}\diagup x\in\mathbb{R}_{>0},~z\in\mathbb{R}\text{, }%
\theta\in S^{1}\right\}  \label{2d-sympl-leaf}%
\end{equation}
\textit{where }$X_{\theta}=\cos\theta X_{1}+\sin\theta X_{2}$.
\end{description}

The zero dimensional orbits lack of interest for our purpose, so let us focus
our attention on the 2-dimensional ones. They are semiplanes orthogonal to the
plane $X_{1},X_{2}$, spanned radially from the $X_{3}$ axis like the pages of
a book, without touching it, and characterized by the angle $\theta$ between
the $X_{1}$ axis and the intersection of the leaf with the $X_{1},X_{2}$ plane.

To write out the explicit form of the $B$ action on $\mathfrak{su}_{2}$, we
parametrize an arbitrary element $\tilde{b}\in B$ as%
\begin{equation}
\tilde{b}=\left[
\begin{array}
[c]{cc}%
a & b+ic\\
0 & a^{-1}%
\end{array}
\right]  \label{b in B}%
\end{equation}
with $a\in\mathbb{R}_{>0}$ and $b,c\in\mathbb{R}$. Then, $\psi\circ
\mathrm{Ad}_{\tilde{b}^{-1}}^{\ast}\circ\psi^{-1}:\mathfrak{su}_{2}%
\longrightarrow\mathfrak{su}_{2}$, in the basis $\left\{  X_{1},X_{2}%
,X_{3}\right\}  \subset\mathfrak{su}_{2}$ gives
\begin{equation}%
\begin{array}
[c]{l}%
(\psi\circ\left(  \mathrm{Ad}_{\tilde{b}^{-1}}\right)  ^{\ast}\circ\psi
^{-1})X_{1}=\dfrac{b}{a}X_{3}+a^{-2}X_{1}\\
\\
(\psi\circ\left(  \mathrm{Ad}_{\tilde{b}^{-1}}\right)  ^{\ast}\circ\psi
^{-1})X_{2}=-\dfrac{c}{a}X_{3}+a^{-2}X_{2}\\
\\
(\psi\circ\left(  \mathrm{Ad}_{\tilde{b}^{-1}}\right)  ^{\ast}\circ\psi
^{-1})X_{3}=X_{3}%
\end{array}
\label{B action on g}%
\end{equation}
so that on the orbit $\mathcal{O}_{\theta}$ it acts as%
\[
(\psi\circ\left(  \mathrm{Ad}_{\tilde{b}^{-1}}\right)  ^{\ast}\circ\psi
^{-1})\left(  xX_{\theta}+zX_{3}\right)  =xa^{-2}X_{\theta}+\left(  z+\frac
{x}{a}\left(  b\cos\theta-c\sin\theta\right)  \right)  X_{3}%
\]
Hence, the \emph{stabilizer} of $X\in\mathcal{O}_{\theta}$ is the \emph{normal
subgroup }$B_{\theta}\subset B$ composed by the matrices
\begin{equation}
\tilde{b}_{\theta}:=%
\begin{pmatrix}
1 & d\left(  \sin\theta+i\cos\theta\right) \\
0 & 1
\end{pmatrix}
\label{stab matrix}%
\end{equation}
with $d\in\mathbb{R}$. The Lie algebra $\mathfrak{b}_{\theta}$ is generated by
the element%
\[
E_{\theta}=\sin\theta~E+\cos\theta~\left(  iE\right)
\]
and, consequently, $\mathfrak{b}/\mathfrak{b}_{\theta}$ is spanned by the
images in the quotient of the elements%

\[%
\begin{array}
[c]{ccc}%
H=%
\begin{pmatrix}
1 & 0\\
0 & -1
\end{pmatrix}
& , & \tilde{E}_{\theta}=%
\begin{pmatrix}
0 & \left(  \cos\theta-i\sin\theta\right) \\
0 & 0
\end{pmatrix}
=\cos\theta~E-\sin\theta~\left(  iE\right)
\end{array}
\]

\subsection{Orbits and Bruhat decomposition\label{orbits and bruhat}}

Let us now describe an issue which will be of central importance in defining
the dualizable subspaces in cotangent bundle of the compact factor $K$. As it
was mentioned in $\left(  \ref{dress action}\right)  $, the action of the
solvable factor $B$ on this phase space arises from the lift of the dressing
action and its orbits on $K$ are the dressing orbits.

The dressing orbits of the Poisson-Lie structure associated to the Iwasawa
decomposition \cite{Lu-We} in a semisimple group can be described by using the
\emph{Bruhat decomposition} \cite{ChPr,Hel}. Let us begin with a compact Lie
group $K$; let $G$ be its complexification. For $G=KB$, the Iwasawa
decomposition associated to $K$, let us choose in the Lie algebra
$\mathfrak{k}$ a maximal abelian subalgebra $\mathfrak{t}$; then
$\mathfrak{h}:=\mathfrak{t}+i\mathfrak{t}$ is a Cartan subalgebra for
$\mathfrak{g}$. Let us fix some ordering of the roots associated to
$\mathfrak{h}$. For example, if $K=SU\left(  n\right)  $ then $G=SL\left(
n,\mathbb{C}\right)  $ and we can choose the order in the roots such that $B$
is the set of upper triangular matrices with real diagonal entries. Let
$T\subset K$ be the connected subgroup associated to $\mathfrak{t}$.

\begin{description}
\item[Lemma:] \textit{The set}
\[
T\cdot B:=\left\{  tb:t\in T,b\in B\right\}
\]
\textit{is a Lie subgroup of }$G$\textit{; moreover, we have that }$T\cdot
B=B\cdot T$\textit{.}
\end{description}

\textbf{Proof: } Because $tBt^{-1}\subset B$ for all $t\in T$, we have that
$T\cdot B$ is a subgroup of $G$ and $T\cdot B=B\cdot T$; if $\left(
c_{n}\right)  \subset T\cdot B$ is a sequence convergent in $G$, we have
sequences $\left(  a_{n}\right)  \subset T,\left(  b_{n}\right)  \subset B$
such that $c_{n}=a_{n}b_{n}$ for all $n\in\mathbb{N}$. Now, because $T$ is
compact, there exists a convergent subsequence $\left(  a_{n_{k}}\right)  $,
and $a_{n_{k}}\to a\in T$. Thus the sequence $b_{n_{k}}=\left(  a_{n_{k}%
}\right)  ^{-1}c_{n_{k}}$ has all its terms in $B$, and it is convergent in
$G$, due to the continuity of the group operations. But $B$ is closed in $G$,
thus $b_{n_{k}}\to b\in B$. Therefore $c_{n}\to ab\in T\cdot B$, and $T\cdot
B$ is a closed subgroup in $G$. $\blacksquare$\newline In the example
considered above, $T$ is the set composed of diagonal matrices whose
nonvanishing entries are elements of $S^{1}$; then $B_{+}:=B\cdot T$ is the
group of upper triangular matrices. Let $N\left(  T\right)  $ the
\emph{normalizer} of $T$: It consists of the elements $k\in K$ such that
$kTk^{-1}\subset T$; then the group $W:=N\left(  T\right)  /T$ is the
\emph{Weyl group} of $K$.

\begin{description}
\item[Theorem (Bruhat decomposition):] \textit{The group }$G$\textit{ can be
decomposed as}
\[
G=\coprod_{w\in W}B_{+}wB_{+}.
\]

\end{description}

In order to use this decomposition, a set of representatives must be chosen
for the elements of $W$. For example, $W$ for $SU\left(  n\right)  $ is the
set of permutations of $n$ elements, and representatives for two-cycles are
\[
s_{i}:=\left[
\begin{array}
[c]{cccccccc}%
1 & 0 & \cdots & 0 & 0 & \cdots & 0 & 0\\
0 & 1 & \cdots & 0 & 0 & \cdots & 0 & 0\\
\vdots & \vdots & \ddots & \vdots & \vdots &  & \vdots & \vdots\\
0 & 0 & \cdots & 0 & 1 & \cdots & 0 & 0\\
0 & 0 & \cdots & -1 & 0 & \cdots & 0 & 0\\
\vdots & \vdots &  & \vdots & \vdots & \ddots & \vdots & \vdots\\
0 & 0 & \cdots & 0 & 0 & \cdots & 1 & 0\\
0 & 0 & \cdots & 0 & 0 & \cdots & 0 & 1
\end{array}
\right]
\]
with the permutation matrix in the $i,i+1$-entries. The disjoint sets in the
Bruhat decomposition gives a kind of cellular decomposition with a unique open
cell plus lower dimensional submanifolds. In the $SU\left(  2\right)  $ case,
representatives for the Weyl groups members are the identity matrix and the
element
\[
\sigma:=\left[
\begin{array}
[c]{cc}%
0 & 1\\
-1 & 0
\end{array}
\right]
\]
so the open submanifold is the set
\[
B_{+}\sigma B_{+}=\left\{  \left[
\begin{array}
[c]{cc}%
a & b\\
c & d
\end{array}
\right]  \in SL\left(  2,\mathbb{C}\right)  :c\not =0\right\}  .
\]
Note that the Bruhat decomposition yields to the decomposition
\[
G=\coprod_{%
\begin{array}
[c]{c}%
w\in W\\
t\in T
\end{array}
}tBwB
\]
by using the fact that $wTw^{-1}\subset T$ for every $w\in N\left(  T\right)
$. On $SL\left(  2,\mathbb{C}\right)  $ this decomposition can be written as
\[
SL\left(  2,\mathbb{C}\right)  =\left(  \coprod_{t\in T}t\cdot B\right)
\amalg\left(  \coprod_{t\in T}t\cdot B^{w}\right)
\]
where $B^{w}$ is the subset of $SL\left(  2,\mathbb{C}\right)  $ composed of
those matrices with its lower-left element strictly negative.

By definition, the dressing orbits in $K$ are the sets $\pi_{K}\left(
Bk\right)  $ for $k\in K$. With the previous decomposition at hands, it is
possible to characterize the orbits of the $B$-action on $K$: In fact, if
$w\in W$, let us denote by $\Sigma_{w}$ the $B$-orbit through $w$: $\Sigma
_{w}=\pi_{K}\left(  Bw\right)  $ (by fixing a set of representatives in $K$
for the element $w\in W$). Then we have the following result.

\begin{description}
\item[Proposition:] \textit{The orbits of the }$B$-\textit{action on }%
$K$\textit{ can be parametrized by }$T\times W$: \emph{That is, every orbit
can be written as }$t\cdot\Sigma_{w}$ \textit{for some }$\left(  w,t\right)
\in W\times T$.
\end{description}

\textbf{Proof: }Let us denote by $\mathcal{O}_{k}$ the $B$-orbit through $k\in
K$; then by using the previous decomposition we can write $k=tb_{1}wb_{2}$ for
some $b_{1},b_{2}\in B,t\in T$ and $w\in W$. Therefore
\[
\mathcal{O}_{k}=\pi_{K}\left(  Bk\right)  =\pi_{K}\left(  Btb_{1}w\right)
=\pi_{K}\left(  tBw\right)  =t\pi_{K}\left(  Bw\right)  =t\cdot\Sigma_{w}%
\]
where it was used that $tB=Bt$. $\blacksquare$\newline

In the case $G=SL\left(  2,\mathbb{R}\right)  $, the orbits are
\[
\Sigma_{1}:=\pi_{SU\left(  2\right)  }\left(  B\cdot\text{id}\right)
=\left\{  \text{id}\right\}
\]
and
\[
\Sigma_{-1}:=\pi_{SU\left(  2\right)  }\left(  B\left[
\begin{array}
[c]{cc}%
0 & 1\\
-1 & 0
\end{array}
\right]  \right)  =\left\{  \left[
\begin{array}
[c]{cc}%
\alpha & b\\
-b & \bar{\alpha}%
\end{array}
\right]  :\alpha\in\mathbb{C},b\in\mathbb{R}^{+},\left\vert \alpha\right\vert
^{2}+b^{2}=1\right\}
\]
So the orbits of the $B$-action on $SU\left(  2\right)  $ are the
zero-dimensional ones
\[
t\cdot\Sigma_{1}=\left\{  \left[
\begin{array}
[c]{cc}%
t & 0\\
0 & t^{-1}%
\end{array}
\right]  \right\}
\]
and the two-dimensional, given by
\[
t\cdot\Sigma_{-1}=\left\{  \left[
\begin{array}
[c]{cc}%
\alpha & \beta\\
-\bar{\beta} & \bar{\alpha}%
\end{array}
\right]  :\arg\beta=\arg t+\pi\right\}
\]
A comment on the choice of representatives for the elements of $W$ is in
order: Any other choice just gives another parametrization $W\times
K\rightarrow\left\{  \mathcal{O}_{k}:k\in K\right\}  $.

\section{Hamiltonian $B$-spaces}

In this section we describe some hamiltonian $B$-spaces related to the
$2$-dimensional symplectic leaves $\mathcal{O}_{\theta}$ $\left(
\ref{2d-sympl-leaf}\right)  $ which in turn will assemble the $T$-duality scheme.

\subsection{2-dimensional symplectic leaves $\mathcal{O}_{\theta}%
\subset\mathfrak{su}_{2}$}

The semiplanes $\mathcal{O}_{\theta}\subset\mathfrak{su}_{2}$ turn in
symplectic manifolds when endowed with the pullback by $\psi:\mathfrak{su}%
_{2}\rightarrow\mathfrak{b}^{\ast}$ of the Kirillov-Kostant structure on the
coadjoint orbits in $\mathfrak{b}^{\ast}$%
\[
\left\langle \omega_{\mathcal{O}_{\theta}},\bar{\psi}_{\ast}\mathrm{ad}%
_{X}^{\ast}\psi\left(  Z\right)  \otimes\bar{\psi}_{\ast}\mathrm{ad}_{Y}%
^{\ast}\psi\left(  Z\right)  \right\rangle =\left(  Z,\left[  X,Y\right]
\right)  _{\mathfrak{su}_{2}}%
\]
where $\bar{\psi}:\mathfrak{b}^{\ast}\rightarrow\mathfrak{su}_{2}$ is the
inverse mapping of $\psi$. They also are hamiltonian $B$-spaces under the
action $\left(  \ref{B action on g}\right)  $.

\subsection{$\mathbb{R}^{2}$ as a phase space}

Given the phase space $\mathbb{R}^{2}$ with coordinates $\left(  q,p\right)
$, there exist a family of embeddings which can be interpreted as the momentum
map associated with some action of $B$ on $\mathbb{R}^{2}$, as explained in
the following proposition.

\begin{description}
\item[Proposition:] \textit{The maps }$\rho:B\times\mathbb{R}^{2}%
\longrightarrow\mathbb{R}^{2}$\textit{ defined as }
\begin{equation}
\rho\left(  \tilde{b},(q,p)\right)  =\left(  q-\frac{1}{\mu}\ln a,p-2\mu
\dfrac{\varepsilon}{a}\exp(2\mu q)\left(  b\cos\theta-c\sin\theta\right)
\right)  \label{rho 0}%
\end{equation}
\textit{are a family of transitive actions of }$B$ \textit{on }$\mathbb{R}%
^{2}$\textit{, for }$(q,p)\in\mathbb{R}^{2}$\textit{, }$\tilde{b}\in B$
\textit{as given in }$\left(  \ref{b in B}\right)  $, $\theta\in\left[
0,2\pi\right]  $\textit{, }$\mu\in\mathbb{R}_{>0}$\textit{ and }%
$\varepsilon\in\mathbb{R}$\textit{,} \textit{arbitrary parameters}.
\textit{Moreover, regarding }$\mathbb{R}^{2}$ \textit{as a symplectic space
endowed with the canonical} \textit{symplectic form} $\omega_{\circ}=dq\wedge
dp$\textit{,} $\left(  \mathbb{R}^{2},\omega_{\circ}\right)  $\textit{, it}
\textit{becomes in an homogeneous hamiltonian }$B$\textit{-space with
associated equivariant momentum map }$\sigma_{\theta}:\mathbb{R}%
^{2}\hookrightarrow\mathfrak{b}^{\ast}$
\[
\sigma_{\theta}(q,p)=-\dfrac{1}{\mu}p\mathbf{h}+\varepsilon\exp(2\mu
q)(\cos\theta\mathbf{e}-\sin\theta\mathbf{\tilde{e}})
\]
\textit{For each fixed value of }$\theta$\textit{, the induced map }%
$\tilde{\sigma}_{\theta}:\mathbb{R}^{2}\hookrightarrow\mathfrak{su}_{2}$
\[
\tilde{\sigma}_{\theta}(q,p):=\psi^{-1}\circ\sigma_{\theta}\left(  q,p\right)
=\dfrac{1}{2\mu}pX_{3}-\varepsilon\exp(2\mu q)(\cos\theta X_{1}+\sin\theta
X_{2})
\]
\textit{is a symplectic isomorphism between }$\left(  \mathbb{R}^{2}%
,\omega_{\circ}\right)  $ and $\left(  \mathcal{O}_{\theta},\psi^{\ast}%
\omega_{KK}\right)  $\textit{, where }$\omega_{KK}$ \textit{is the
Kirillov-Kostant symplectic form.}
\end{description}

\textbf{Proof: }It is straightforward to check that $\rho$ is a transitive
action and that it is hamiltonian. The infinitesimal generator associated to
$\tilde{X}\in\mathfrak{b}$,%
\[
\tilde{X}=uE+v\left(  iE\right)  +wH=\left[
\begin{array}
[c]{cc}%
w & u+iv\\
0 & -w
\end{array}
\right]
\]
can be calculated from the expression%
\[
e^{t\tilde{X}}=\left[
\begin{array}
[c]{cc}%
e^{tw} & \frac{1}{w}\sinh\left(  tw\right)  \left(  u+iv\right) \\
0 & e^{-tw}%
\end{array}
\right]
\]
giving%
\[
\left.  \tilde{X}_{\mathbb{R}^{2}}\right\vert _{(q,p)}=(-\dfrac{1}{\mu
}w,-2\varepsilon\mu\exp(2\mu q)\left(  u\cos\theta-v\sin\theta\right)  )
\]
The contraction of this vector with the symplectic form is
\[
\imath_{\tilde{X}_{\mathbb{R}^{2}}}\left(  dq\wedge dp\right)  =d\left\langle
-\dfrac{1}{\mu}p\mathbf{h}+\varepsilon\exp(2\mu q)(\cos\theta\mathbf{e}%
-\sin\theta\mathbf{\tilde{e}}),\tilde{X}\right\rangle
\]
from where we get the momentum map $\sigma_{\theta}(q,p)$.

The last statement is a direct consequence of the equivariance property. Under
the action $\psi\circ\mathrm{Ad}_{\tilde{b}^{-1}}^{\ast}\circ\psi
^{-1}:\mathfrak{su}_{2}\longrightarrow\mathfrak{su}_{2}$ it behaves as
\begin{align*}
\left(  \psi\circ\mathrm{Ad}_{\tilde{b}^{-1}}^{\ast}\circ\psi^{-1}\right)
\tilde{\sigma}(q,p)=  &  -\dfrac{1}{2}\left(  p+\frac{2\varepsilon\exp(2\mu
q)}{a}\left(  b\cos\theta-c\sin\theta\right)  \right)  X_{3}\\
&  -\frac{\varepsilon\exp{(}2\mu q{)}}{a^{2}}(\cos\theta~X_{1}+\sin
\theta~X_{2})
\end{align*}
satisfying the equivariant property $\tilde{\sigma}_{\theta}(\rho_{\tilde{b}%
}(q,p))=\left(  \psi\circ\mathrm{Ad}_{\tilde{b}^{-1}}^{\ast}\circ\psi
^{-1}\right)  \tilde{\sigma}(q,p)$.$\blacksquare$

\subsection{The cotangent bundle of $B$}

Let us consider the phase space $T^{\ast}B=B\times\mathfrak{b}^{\ast}$,
trivialized by left translation, endowed with the canonical symplectic form
$\tilde{\omega}_{\circ}$. It is a hamiltonian $B$-space by the hamiltonian
action of $B$ on $B\times\mathfrak{b}^{\ast}$
\[
\lambda:B\times\left(  B\times\mathfrak{b}^{\ast}\right)  \longrightarrow
\left(  B\times\mathfrak{b}^{\ast}\right)  ~\diagup~\lambda\left(  \tilde
{h},\left(  \tilde{b},\tilde{\eta}\right)  \right)  =\left(  \tilde{h}%
\tilde{b},\tilde{\eta}\right)
\]
for $\tilde{h},\tilde{b}\in B$, $X\in\mathfrak{b}^{\ast}$, with associated
momentum map $\mu:B\times\mathfrak{b}^{\ast}\longrightarrow\mathfrak{b}^{\ast
}~/$ $\mu\left(  \tilde{b},\tilde{\eta}\right)  =\mathrm{Ad}_{\tilde{b}^{-1}%
}^{\ast}\tilde{\eta}$. The corresponding map $\tilde{\mu}:B\times
\mathfrak{b}^{\ast}\longrightarrow\mathfrak{su}_{2}~$with $\tilde{\mu}%
=\psi^{-1}\circ\mu$ is
\[
\tilde{\mu}\left(  \tilde{b},\tilde{\eta}\right)  =\left(  \bar{\psi}%
\circ\mathrm{Ad}_{\tilde{b}^{-1}}^{\ast}\circ\psi\right)  \psi\left(
\tilde{\eta}\right)
\]
that has the explicit form%
\begin{equation}
\tilde{\mu}\left(  \tilde{b},\tilde{\eta}_{\mathbf{e}}\mathbf{e}+\tilde{\eta
}_{\mathbf{\tilde{e}}}\mathbf{\tilde{e}}+\tilde{\eta}_{\mathbf{h}}%
\mathbf{h}\right)  =-a^{-2}\tilde{\eta}_{\mathbf{e}}X_{1}+a^{-2}\tilde{\eta
}_{\mathbf{\tilde{e}}}X_{2}-\left(  \frac{1}{2}\tilde{\eta}_{\mathbf{h}%
}+\dfrac{b\tilde{\eta}_{\mathbf{e}}+c\tilde{\eta}_{\mathbf{\tilde{e}}}}%
{a}\right)  X_{3} \label{m-map in T*B}%
\end{equation}
where we parametrized an element $\tilde{b}\in B$ as%
\[
\tilde{b}=\left(
\begin{array}
[c]{cc}%
a & b+ic\\
0 & -a
\end{array}
\right)
\]
with $a\in\mathbb{R}^{+}$, $b,c\in\mathbb{R}$.

\subsection{The cotangent bundle of $SU\left(  2\right)  $}

The third phase space we consider here is the cotangent bundle of the
remaining factor of the factorization of $SL(2,\mathbb{C})$, namely $T^{\ast
}SU\left(  2\right)  $. To stand the notation to be used in rest of this work,
we parametrize an element $g\in SU\left(  2\right)  $ as
\[
g=\left(
\begin{array}
[c]{cc}%
\alpha & \beta\\
-\bar{\beta} & \bar{\alpha}%
\end{array}
\right)
\]
where the bar over the complex entries $\alpha,\beta$ is meant to indicate the
complex conjugate. We regard $T^{\ast}SU\left(  2\right)  $ trivialized by
left translation, $T^{\ast}SU\left(  2\right)  =SU\left(  2\right)
\times\mathfrak{su}_{2}^{\ast}$, endowed with the canonical symplectic form
$\omega_{\circ}$. In order to turn it into a hamiltonian $B$-space, we are
given the dressing action $\mathsf{d}:B\times SU\left(  2\right)
\longrightarrow SU\left(  2\right)  $ which arises from the factorization
$SL(2,\mathbb{C})=SU\left(  2\right)  \times B$ such that, for $\tilde{b}\in
B$ and $g\in SU\left(  2\right)  $, $\mathsf{Dr}\left(  \tilde{b},g\right)
=\Pi_{SU\left(  2\right)  }\left(  \tilde{b}g\right)  =g^{\tilde{b}}$. It is
lifted to $SU\left(  2\right)  \times\mathfrak{su}_{2}^{\ast}$ as explained in
the following proposition.

\begin{description}
\item[Theorem:] \textit{The action }$\mathsf{d}:B\times SU\left(  2\right)
\longrightarrow SU\left(  2\right)  $, \textit{defined above as }%
$\mathsf{d}_{\tilde{b}}\left(  g\right)  =g^{\tilde{b}}$\textit{,}
\textit{lift to the cotangent bundle in body coordinates, }$SU\left(
2\right)  \times\mathfrak{su}_{2}^{\ast}$\textit{, as}%
\begin{equation}
\mathsf{\hat{d}}_{\tilde{b}}\left(  g,\eta\right)  =\left(  g^{\tilde{b}%
},\left(  \psi^{\ast}\circ\mathrm{Ad}_{\tilde{b}^{g}}\circ\bar{\psi}^{\ast
}\right)  \eta\right)  \label{dress action}%
\end{equation}
\textit{It is a symplectic action with }$\mathrm{Ad}$\textit{-equivariant
momentum map }$\varphi:SU\left(  2\right)  \times\mathfrak{su}_{2}^{\ast
}\longrightarrow\mathfrak{b}^{\ast}$%
\begin{equation}
\varphi(g,\eta)=\psi\left(  \Pi_{\mathfrak{su}_{2}}\left(  \mathrm{Ad}_{g}%
^{G}\bar{\psi}^{\ast}\left(  \eta\right)  \right)  \right)
\label{momentum map 2}%
\end{equation}
\textit{where }$\bar{\psi}^{\ast}:\mathfrak{su}_{2}^{\ast}\longrightarrow
\mathfrak{b}$ \textit{is the pullback of the bijection }$\mathfrak{b}^{\ast
}\overset{\bar{\psi}}{\longrightarrow}\mathfrak{su}_{2}$.
\end{description}

\textbf{Proof: }We get the action on the left trivialized cotangent bundle
from the relation
\[
\left\langle \left(  g^{\tilde{b}},\eta\right)  ,\left(  \mathsf{d}_{\tilde
{b}}\right)  _{\ast B}\left(  g,X\right)  \right\rangle =\left\langle
\eta,X^{\tilde{b}^{g}}\right\rangle
\]
where $X=$ $g^{-1}\dot{g}$ and%
\[
X^{\tilde{b}}=\left.  \frac{d\left(  g^{-1}g\left(  t\right)  \right)
^{\tilde{b}}}{dt}\right\vert _{t=0}=Ad_{\tilde{b}^{g}}^{\ast}X
\]
by using the relations $\left(  gh\right)  ^{\tilde{b}}=g^{\tilde{b}}%
h^{\tilde{b}^{g}}$.

Then, the infinitesimal generator $\left.  \tilde{Z}_{SU\left(  2\right)
\times\mathfrak{su}_{2}^{\ast}}\right\vert _{\left(  g,\eta\right)  }=\left.
\frac{d}{dt}\mathsf{d}_{e^{t\tilde{Z}}}^{SU\left(  2\right)  \times
\mathfrak{su}_{2}^{\ast}}\left(  g,\eta\right)  \right\vert _{t=0}$ is, for
$\tilde{Z}\in\mathfrak{b}$,%
\[
\left.  \tilde{Z}_{SU\left(  2\right)  \times\mathfrak{su}_{2}^{\ast}%
}\right\vert _{\left(  g,\eta\right)  }=\left(  g^{\tilde{Z}},\psi^{\ast
}\left(  \left[  \mathrm{Ad}_{g}^{\ast}\tilde{Z},\bar{\psi}^{\ast}\left(
\eta\right)  \right]  \right)  \right)
\]
where $g^{\tilde{Z}}=\left.  d\left(  g^{e^{t\tilde{Z}}}\right)  \diagup
dt\right\vert _{t=0}$. From this vector field on $SU\left(  2\right)
\times\mathfrak{su}_{2}^{\ast}$ we compute the momentum map since
$\left\langle \eta,g^{-1}g^{\tilde{Z}}\right\rangle =\left\langle
\varphi(g,\eta),\tilde{Z}\right\rangle $, using that $\Pi_{\mathfrak{su}_{2}%
}\mathrm{Ad}_{g^{-1}}^{G}\tilde{Z}=g^{-1}g^{\tilde{Z}}$ and the bijection
$\psi:\mathfrak{su}_{2}\longrightarrow\mathfrak{b}^{\ast}$, its the adjoint
$\psi^{\ast}:\mathfrak{b}\longrightarrow\mathfrak{su}_{2}^{\ast}$ and its
inverse $\bar{\psi}^{\ast}:\mathfrak{su}_{2}^{\ast}\longrightarrow
\mathfrak{b}$, obtaining
\[
\left\langle \varphi(g,\eta),\tilde{Z}\right\rangle _{\mathfrak{g}%
}=\left\langle \psi\left(  \Pi_{\mathfrak{su}_{2}}\mathrm{Ad}_{g}^{G}\bar
{\psi}^{\ast}\left(  \eta\right)  \right)  ,\tilde{Z}\right\rangle
\]
Since it arises as the lifting of a symmetry on the base space $SU\left(
2\right)  $, it is naturally \emph{equivariant}.$\blacksquare$

As in the previous sections, we shall consider momentum maps valued on
$\mathfrak{su}_{2}$, so we define%
\begin{equation}%
\begin{array}
[c]{c}%
\tilde{\varphi}\cong\bar{\psi}\circ\varphi:SU\left(  2\right)  \times
\mathfrak{su}_{2}^{\ast}\longrightarrow\mathfrak{su}_{2}\\
\\
\tilde{\varphi}(g,\eta)=\Pi_{\mathfrak{su}_{2}}\left(  \mathrm{Ad}_{g}^{G}%
\bar{\psi}^{\ast}\left(  \eta\right)  \right)  =g\left(  g^{-1}\right)
^{\bar{\psi}^{\ast}\left(  \eta\right)  }%
\end{array}
\label{dress and mom map}%
\end{equation}
where we used that $\mathrm{Ad}_{g}^{G}\bar{\psi}^{\ast}\left(  \eta\right)
=g\left(  g^{-1}\right)  ^{\bar{\psi}^{\ast}\left(  \eta\right)  }+\bar{\psi
}^{\ast}\left(  \mathrm{Ad}_{g^{-1}}^{\ast}\eta\right)  $. Observe that the
momentum map associated with the dressing action is the Maurer-Cartan form
applied to the infinitesimal generator at each point.

\section{T-duality}

The T-duality scheme involves the three hamiltonian $B$-spaces described
above, linking them with $\mathfrak{su}_{2}\overset{\psi}{\cong}%
\mathfrak{b}^{\ast}$ by equivariant arrows%
\[
\begin{diagram} \node{T^{\ast}SU(2)}\arrow{se,b}{{\tilde\varphi}}\node{\mathbb{R}^2}\arrow{s,r}{\tilde\sigma}\node{T^{\ast}B}\arrow{sw,r}{{\tilde\mu}}\\ \node[2]{{\mathfrak{su}}_{2}} \end{diagram}
\]
It is worth to remark that T-duality is not symplectic equivalence on the full
phase space. Indeed, each symplectic equivalence class is defined by a
coadjoint orbit $\mathcal{O}_{\theta}$ $\left(  \ref{2d-sympl-leaf}\right)  $
and its elements are some symplectic submanifolds contained in $SU\left(
2\right)  \times\mathfrak{su}_{2}^{\ast}$ and $B\times\mathfrak{b}^{\ast}$,
which are called \emph{dualizable subspaces}. They can be defined as the
leaves of some foliation in the pre-images of $\mathcal{O}_{\theta}$ through
the maps $\tilde{\mu}$, $\tilde{\varphi}$, $\tilde{\sigma}_{\theta}$, as it
will be explained below.

Let us consider the three fibrations on $\mathcal{O}_{\theta}$
\[
\begin{diagram} \node{\tilde{\varphi}^{-1}\left( \mathcal{O}_{\theta }\right)}\arrow{se,b}{{\tilde\varphi}}\node{\mathbb{R}^2}\arrow{s,r}{\tilde\sigma}\node{\tilde{\mu}^{-1}\left( \mathcal{O}_{\theta }\right)}\arrow{sw,r}{{\tilde\mu}}\\ \node[2]{\mathcal{O}_{\theta}}\arrow{e,J}\node{{\mathfrak{su}}_{2}} \end{diagram}
\]
where $\tilde{\mu}^{-1}\left(  \mathcal{O}_{\theta}\right)  \subset
B\times\mathfrak{b}^{\ast}$ and $\tilde{\varphi}^{-1}\left(  \mathcal{O}%
_{\theta}\right)  \subset SU\left(  2\right)  \times\mathfrak{su}_{2}^{\ast}$
are \emph{coisotropic submanifolds}, and $\tilde{\sigma}_{\theta}^{-1}\left(
\mathcal{O}_{\theta}\right)  \cong\mathbb{R}^{2}$ is a symplectic space. Let
us take a closer look of them. The tangent spaces of the fibers are the
kernels of the corresponding differential momentum map, and their symplectic
orthogonal are the tangent spaces to the orbits of $B$ through each point.
Then, collective hamiltonians on $B\times\mathfrak{b}^{\ast}$, $\ SU\left(
2\right)  \times\mathfrak{su}_{2}^{\ast}$ and $\mathbb{R}^{2}$ furnish the
compatible dynamics having hamiltonian vector fields tangent to the
$B$-orbits. The equivariant momentum maps carry them over a hamiltonian vector
field tangent to the coadjoint orbit $\mathcal{O}_{\theta}$. This is the main
idea underlying T-duality, establishing a correspondence between hamiltonian
vector fields, so the correspondence between integral curves is defined up to
a shifting of the initial condition.

Let us work out the \emph{dualizable space }in each case.

\subsection{Dualizable subspaces in $B\times\mathfrak{b}^{\ast}$}

Let us denote by $\bar{\tau}_{B}:B\times\mathfrak{b}^{\ast}\rightarrow B$ to
the canonical projection. Sizing up the set obtained by the intersection
between $\tilde{\mu}^{-1}\left(  \mathcal{O}_{\theta}\right)  $ and the fibre
$\bar{\tau}_{B}^{-1}\left(  \tilde{b}\right)  $%
\[
\mu_{1}^{-1}\left(  \mathcal{O}_{\theta}\right)  \cap\bar{\tau}_{B}%
^{-1}\left(  \tilde{b}\right)  =\left\{  \left(  \tilde{b},\eta\right)
\in\left\{  \tilde{b}\right\}  \times\mathfrak{b}^{\ast}:\bar{\psi}\left(
\mathrm{Ad}_{\tilde{b}^{-1}}^{\ast}\tilde{\eta}\right)  \in\mathcal{O}%
_{\theta}\right\}
\]
one realizes that $\tilde{\mu}^{-1}\left(  \mathcal{O}_{\theta}\right)
=B\times\psi\left(  \mathcal{O}_{\theta}\right)  $. It is a coisotropic
submanifold and the null distribution of the presymplectic form $\left.
\tilde{\omega}_{\circ}\right\vert _{\tilde{\mu}^{-1}\mathcal{O}_{\theta}}$ is
spanned by the infinitesimal generators associated to the Lie algebra of the
\emph{stabilizer} subgroup $B_{\theta}$. A more precise description of this
set is
\begin{equation}
\tilde{\mu}^{-1}\mathcal{O}_{\theta}=\left\{  \left(  \tilde{b},\tilde{\eta
}_{\mathbf{h}}\mathbf{h}+\tilde{\eta}_{\mathbf{\tilde{e}}_{\theta}%
}\mathbf{\tilde{e}}_{\theta}\right)  \diagup\tilde{b}\in B,~\tilde{\eta
}_{\mathbf{\tilde{e}}_{\theta}}\in\mathbb{R}^{+},~\tilde{\eta}_{\mathbf{h}}%
\in\mathbb{R}\right\}  = B\times\psi\left(  \mathcal{O}_{\theta}\right)
\label{mu^(-1) O_(theta)}%
\end{equation}
where we introduced the dual basis $\left\{  \mathbf{e}_{\theta}%
,\mathbf{\tilde{e}}_{\theta},\mathbf{h}\right\}  \subset\mathfrak{b}^{\ast}$
with%
\[%
\begin{array}
[c]{ccc}%
\mathbf{\tilde{e}}_{\theta}=\cos\theta~\mathbf{e}-\sin\theta~\mathbf{\tilde
{e}} & , & \mathbf{e}_{\theta}=\sin\theta~\mathbf{e+}\cos\theta~\mathbf{\tilde
{e}}%
\end{array}
\]
Observe that $\mathbf{e}_{\theta}=-\psi\left(  X_{\theta}\right)  $.

In order to determine the presymplectic form on $\tilde{\mu}^{-1}%
\mathcal{O}_{\theta}$, we left-trivialize the canonical vector bundles on $B$;
thus we have the identifications
\[%
\begin{array}
[c]{ccc}%
T^{\ast}T^{\ast}B\simeq\underbrace{B\times\mathfrak{b}^{\ast}}_{\text{base}%
}\times\mathfrak{b}^{\ast}\times\mathfrak{b} &  & TT^{\ast}B\simeq
\overbrace{B\times\mathfrak{b}^{\ast}}^{\text{base}}\times\mathfrak{b}%
\times\mathfrak{b}^{\ast}%
\end{array}
\]
and it yields to the following expression for the canonical $2$-form on
$T^{\ast}T^{\ast}B$
\[
\left.  \omega_{\circ}\right\vert _{\left(  \tilde{b},\tilde{\eta}\right)
}\left(  \left(  \xi_{1},\rho_{1}\right)  ,\left(  \xi_{2},\rho_{2}\right)
\right)  =-\rho_{1}\left(  \xi_{2}\right)  +\rho_{2}\left(  \xi_{1}\right)
+\tilde{\eta}\left(  \left[  \xi_{1},\xi_{2}\right]  \right)  .
\]
Having in mind that $\tilde{\mu}^{-1}\mathcal{O}_{\theta}=E_{\theta}%
^{-1}\left(  0\right)  $ and expanding it in the given basis, we can express
the canonical form restricted to this submanifold as
\[
\left.  \tilde{\omega}_{\circ}\right\vert _{\tilde{\mu}^{-1}\mathcal{O}%
_{\theta}}\left(  \tilde{b},\tilde{\eta}\right)  =2\tilde{\eta}%
_{\mathbf{\tilde{e}}_{\theta}}\left(  \tilde{b}^{-1}\mathbf{h}\wedge\tilde
{b}^{-1}\mathbf{\tilde{e}}_{\theta}\right)  -\tilde{E}_{\theta}\wedge\tilde
{b}^{-1}\mathbf{\tilde{e}}_{\theta}-H\wedge\tilde{b}^{-1}\mathbf{h}%
\]
As a map from $T_{\left(  \tilde{b},\tilde{\eta}\right)  }\tilde{\mu}%
^{-1}\mathcal{O}_{\theta}\longrightarrow T_{\left(  \tilde{b},\tilde{\eta
}\right)  }^{\ast}\tilde{\mu}^{-1}\mathcal{O}_{\theta}$, it assigns to a
vector $V=\left(  \tilde{b}\left(  v_{H}H+v_{E_{\theta}}E_{\theta}%
+v_{\tilde{E}_{\theta}}\tilde{E}_{\theta}\right)  ,\tilde{\xi}_{\mathbf{e}%
_{\theta}}\mathbf{e}_{\theta}+\tilde{\xi}_{\mathbf{\tilde{e}}_{\theta}%
}\mathbf{\tilde{e}}_{\theta}+\tilde{\xi}_{\mathbf{h}}\mathbf{h}\right)  $ the
hamiltonian 1-form%
\[
\imath_{V}\left.  \tilde{\omega}_{\circ}\right\vert _{\tilde{\mu}%
^{-1}\mathcal{O}_{\theta}}=\left(  2\tilde{\eta}_{\mathbf{\tilde{e}}_{\theta}%
}v_{H}-\tilde{\xi}_{\mathbf{\tilde{e}}_{\theta}}\right)  \tilde{b}%
^{-1}\mathbf{\tilde{e}}_{\theta}-\left(  2\tilde{\eta}_{\mathbf{\tilde{e}%
}_{\theta}}v_{\tilde{E}_{\theta}}+\tilde{\xi}_{\mathbf{h}}\right)  \tilde
{b}^{-1}\mathbf{h}+v_{\tilde{E}_{\theta}}\tilde{E}_{\theta}+v_{H}H.
\]

\subsubsection{Gauge fixing$\ $and canonical coordinates}

The evolution of the system is contained in the coisotropic submanifold
$\tilde{\mu}^{-1}\left(  \mathcal{O}_{\theta}\right)  \subset B\times
\mathfrak{b}^{\ast}$. Without doing explicit mention of this fact from now on,
we will use in the current section the identification
\[
B\times\mathfrak{b}^{\ast}\overset{\text{id}\times\bar{\psi}}{\simeq}%
B\times\mathfrak{su}_{2}%
\]
As it is known \cite{Guill-Sten}, the leaf of the null foliation through a
point $\left(  \tilde{b},X\right)  $ in $B\times\mathfrak{su}_{2}$ coincides
with the orbit $B_{\tilde{\zeta}}\cdot\left(  \tilde{b},X\right)  $ of the
isotropy group $B_{\tilde{\zeta}}$ of the element $\tilde{\zeta}:=\tilde{\mu
}\left(  \tilde{b},\tilde{\eta}\right)  $. We also know that $B_{\tilde{\zeta
}}=B_{\tilde{\zeta}^{\prime}}$ for every pair of elements $\tilde{\zeta
},\tilde{\zeta}^{\prime}\in\mathcal{O}_{\theta}$ in the same orbit, so we
denote it as $B_{\theta}$ and its elements where described in eq. $\left(
\ref{stab matrix}\right)  $. Therefore the leaves of the null foliation for
the presymplectic structure on $B\times\mathcal{O}_{\theta}$ are the subsets
\[
B_{\theta}\cdot\left(  \left[
\begin{array}
[c]{cc}%
a & b+ic\\
0 & a^{-1}%
\end{array}
\right]  ,X\right)  =\left\{  \left(  \left[
\begin{array}
[c]{cc}%
a & id\left(  b+ic\right)  e^{-i\theta}\\
0 & a^{-1}%
\end{array}
\right]  ,X\right)  :d\in\mathbb{R}\right\}
\]
%The null foliation through the point is spanned by the isotropy
%group of $\mathcal{O}_{\theta}$. But $\left(  \tilde{b};X\right)
%\in\tilde{\mu}^{-1}\mathcal{O}_{\theta}$ iff $X\in
%\mathcal{O}_{\theta}$, and this set will not be a symplectic
%submanifold (it has dimension $5$). The orbits of the $S^{1}$-action%
%\begin{align*}
%&  e^{i\phi}\cdot\left(  \left[
%\begin{array}
%[c]{cc}%
%a & b+ic\\
%0 & a^{-1}%
%\end{array}
%\right]  ;n_{0}\left(  \cos\theta X_{1}+\sin\theta X_{2}\right)  +m_{0}%
%X_{3}\right) \mapsto\\
%&  \mapsto\left(  \left[
%\begin{array}
%[c]{cc}%
%a & \left(  b+ic\right)  e^{i\phi}\\
%0 & a^{-1}%
%\end{array}
%\right]  ;n_{0}\left(  \cos\theta X_{1}+\sin\theta X_{2}\right)  +m_{0}%
%X_{3}\right)
%\end{align*}
%are the leaves of the null foliation of the restriction to $B\times
%\mathcal{O}_{\theta}$ of the symplectic form in $B\times\mathfrak{b}^{\ast}$.
Then any slice for the action of $B_{\theta}$ on $B\times\mathcal{O}_{\theta}$
gives a symplectic submanifold of $B\times\mathfrak{su}_{2}$. As such slice we
can consider the submanifold $\mathcal{S}$ parameterized by $\mathbb{R}%
\times\mathbb{R}^{+}\times\mathbb{R}^{2}$ through the map
\[
\Psi:\left(  t,a;v_{3},v_{\theta}\right)  \mapsto\left(  \left[
\begin{array}
[c]{cc}%
a & -te^{-i\theta}\\
0 & a^{-1}%
\end{array}
\right]  ,-\frac{1}{2}v_{3}X_{3}-v_{\theta}X_{\theta}\right)
\]
where $X_{\theta}:=\cos{\theta}X_{1}+\sin{\theta}X_{2}$. By taking into
account the dual basis $\left\{  \mathbf{x}_{i}\right\}  \subset
\mathfrak{su}_{2}^{\ast}$ and the form $\mathbf{x}_{\theta}:=-\sin{\theta
}\mathbf{x}_{1}+\cos{\theta}\mathbf{x}_{2}$, the constraints for $\mathcal{S}$
are
\[%
\begin{cases}
F_{1}\left(  \tilde{b},X\right)  :=\mathbf{x}_{\theta}\left(  X\right)  =0\\
F_{2}\left(  \tilde{b},X\right)  :=b\sin{\theta}+c\cos{\theta}=0
\end{cases}
\]

\begin{description}
\item[Proposition:] \textit{The pullback of the canonical }$1$-\textit{form
along }$\Psi$\textit{ is given by}
\[
\left.  \left(  \left.  \theta_{0}\right\vert \mathcal{S}\right)  \right\vert
_{\left(  t,a;v_{3},v_{\theta}\right)  }=-a^{-2}\left(  tv_{\theta}%
-v_{3}\right)  da-a^{-1}v_{\theta}dt.
\]
\textit{Then the canonical }$2$-\textit{form on }$\mathcal{S}$\textit{ is }
\[
\omega_{\mathcal{S}}=-a^{-2}\left(  tdv_{\theta}-dv_{3}\right)  \wedge
da+2a^{-2}v_{\theta}da\wedge dt-a^{-1}dv_{\theta}\wedge dt.
\]
\textit{Furthermore it can be written as}
\[
\omega_{\mathcal{S}}=-dp_{a}\wedge da-dp_{t}\wedge dt
\]
\textit{where }$p_{a}\left(  a,t;v_{3},v_{\theta}\right)  :=a^{-2}\left(
tv_{\theta}-v_{3}\right)  ,p_{t}\left(  a,t;v_{3},v_{\theta}\right)
:=a^{-1}v_{\theta}$.
\end{description}

\textbf{Proof:} It is immediate to show that%
\begin{align*}
\left[  \Psi\left(  t,a;v_{3},v_{\theta}\right)  \right]  ^{-1}\Psi_{\ast
}\left(  \frac{\partial}{\partial a}\right)   &  =\left(  \left[
\begin{array}
[c]{cc}%
a^{-1} & -a^{-2}te^{-i\theta}\\
0 & -a^{-1}%
\end{array}
\right]  ;0\right) \\
& \\
\left[  \Psi\left(  t,a;v_{3},v_{\theta}\right)  \right]  ^{-1}\Psi_{\ast
}\left(  \frac{\partial}{\partial t}\right)   &  =\left(  \left[
\begin{array}
[c]{cc}%
0 & -a^{-1}e^{-i\theta}\\
0 & 0
\end{array}
\right]  ;0\right)
\end{align*}
It implies%
\begin{align*}
\left.  \left(  \left.  \theta_{0}\right\vert \mathcal{S}\right)  \right\vert
_{\left(  t,a;v_{3},v_{\theta}\right)  }  &  =\left(  \left[  \Psi\left(
t,a;v_{3},v_{\theta}\right)  \right]  ^{-1}\Psi_{\ast}\left(  \frac{\partial
}{\partial a}\right)  ,-\frac{1}{2}v_{3}X_{3}-v_{\theta}X_{\theta}\right)
_{\mathfrak{sl}\left(  2,\mathbb{C}\right)  }da+\\
&  +\left(  \left[  \Psi\left(  t,a;v_{3},v_{\theta}\right)  \right]
^{-1}\Psi_{\ast}\left(  \frac{\partial}{\partial t}\right)  ,-\frac{1}{2}%
v_{3}X_{3}-v_{\theta}X_{\theta}\right)  _{\mathfrak{sl}\left(  2,\mathbb{C}%
\right)  }dt
\end{align*}
so, we can write
\[
\left(  \left.  \theta_{0}\right\vert \mathcal{S}\right)  =-a^{-2}\left(
t\tilde{E}_{\theta}-aH\right)  da-a^{-1}\tilde{E}_{\theta}dt
\]
The rest follows by exterior differentiation.$\blacksquare$

\begin{description}
\item[Corollary:] \textit{The constraints describing the submanifold
}$\mathcal{S}$\textit{ are of second order.}
\end{description}

\textbf{Proof:} For each pair $\left(  a,t\right)  \in\mathbb{R}^{+}%
\times\mathbb{R}$, the map
\[
\left(  v_{3},v_{\theta}\right)  \mapsto\left(  a^{-2}\left(  tv_{\theta
}-v_{3}\right)  ,a^{-1}v_{\theta}\right)
\]
is non singular, and maps $\omega_{\mathcal{S}}$ on a non degenerate $2$-form,
as the previous proposition shows. Then $\omega_{\mathcal{S}}$ is non
degenerate, and $\mathcal{S}$ is a symplectic submanifold.$\blacksquare$

\subsection{Dualizable subspaces in $SU\left(  2\right)  \times\mathfrak{su}%
_{2}^{\ast}$}

In this case, the \textit{dualizable spaces} are the symplectic leaves of the
coisotropic submanifold $\tilde{\varphi}^{-1}\left(  \mathcal{O}_{\theta
}\right)  $. To get some insight about this set, we consider the infinitesimal
generators of the dressing action of $B$ on $SU(2)$ associated with the basis
$\left\{  E,\left(  iE\right)  ,H\right\}  $ $\subset\mathfrak{b}$%
\begin{equation}%
\begin{array}
[c]{l}%
g^{-1}g^{H}=-i\left(  \alpha\bar{\beta}-\bar{\alpha}\beta\right)
X_{1}-\left(  \alpha\bar{\beta}+\bar{\alpha}\beta\right)  X_{2}\\
g^{-1}g^{\left(  iE\right)  }=-\frac{1}{2}\left(  \beta^{2}+\bar{\beta}%
^{2}\right)  X_{1}-\frac{1}{2}i\left(  \beta^{2}-\bar{\beta}^{2}\right)
X_{2}\mathbf{-}\frac{1}{2}\left(  \bar{\alpha}\bar{\beta}+\alpha\beta\right)
X_{3}\\
g^{-1}g^{E}=-\frac{1}{2}i\left(  \beta^{2}-\bar{\beta}^{2}\right)  X_{1}%
+\frac{1}{2}\left(  \beta^{2}+\bar{\beta}^{2}\right)  X_{2}-\frac{1}%
{2}i\left(  \alpha\beta-\bar{\alpha}\bar{\beta}\right)  X_{3}%
\end{array}
\label{dress inf gen}%
\end{equation}
where $\psi:\mathfrak{su}_{2}\rightarrow\mathfrak{b}^{\ast}$ was given in eq.
$\left(  \ref{psi}\right)  $. It is worth remarking that they can be also
obtained from the coboundary Poisson bivector on $SU(2)$%
\begin{equation}
\pi_{SU(2)}\left(  g\right)  =\frac{1}{4}\left(  gX_{2}\mathbf{\otimes}%
gX_{1}-gX_{1}\mathbf{\otimes}gX_{2}-X_{2}g\mathbf{\otimes}X_{1}g+X_{1}%
g\mathbf{\otimes}X_{2}g\right)  \label{Poisson bivector on SU(2)}%
\end{equation}
as
\[
g^{-1}g^{\bar{\psi}^{\ast}\left(  \eta\right)  }=L_{g^{-1}\ast}\left[  \left(
id\otimes\eta\circ R_{g^{-1}\ast}\right)  \pi_{SU\left(  2\right)  }\left(
g\right)  \right]
\]
so%
\[
\tilde{\varphi}\left(  g,\eta\right)  \equiv g\left(  g^{-1}\right)
^{\bar{\psi}^{\ast}\left(  \eta\right)  }=L_{g\ast}\left[  \left(
id\otimes\eta\circ R_{g\ast}\right)  \pi_{SU\left(  2\right)  }\left(
g^{-1}\right)  \right]
\]

The dual map $\bar{\psi}^{\ast}:\mathfrak{su}_{2}^{\ast}\rightarrow
\mathfrak{b}$ for the dual base $\left\{  \mathbf{x}_{1},\mathbf{x}%
_{2},\mathbf{x}_{3}\right\}  \subset\mathfrak{su}_{2}^{\ast}$ gives%
\[
\bar{\psi}^{\ast}(\mathbf{x}_{1})=-E,\qquad\bar{\psi}^{\ast}(\mathbf{x}%
_{2})=iE,\qquad\bar{\psi}^{\ast}(\mathbf{x}_{3})=-\frac{1}{2}H
\]
which allows us to write%
\[
\bar{\psi}^{\ast}\left(  \eta\right)  =-\eta_{1}E+\eta_{2}\left(  iE\right)
-\frac{1}{2}\eta_{3}H=\left[
\begin{array}
[c]{cc}%
-\frac{1}{2}\eta_{3} & -\eta_{1}+i\eta_{2}\\
0 & \frac{1}{2}\eta_{3}%
\end{array}
\right]
\]
so we get the explicit form of the momentum map $\tilde{\varphi}:SU\left(
2\right)  \times\mathfrak{su}_{2}^{\ast}\longrightarrow\mathfrak{su}_{2}$
\begin{align}
&  g\left(  g^{-1}\right)  ^{\bar{\psi}^{\ast}\left(  \eta\right)
}\nonumber\\
&  =\left(  -\operatorname{Im}\left(  \beta^{2}\right)  \left(  \eta_{1}%
\cos\theta-\eta_{2}\sin\theta\right)  -\operatorname{Re}\left(  \beta
^{2}\right)  \left(  \eta_{1}\sin\theta+\eta_{2}\cos\theta\right)  -\eta
_{3}\operatorname{Im}\left(  \alpha\beta e^{i\theta}\right)  \right)
X_{\theta}\nonumber\\
&  ~~~~~+\left(  \operatorname{Im}\left(  \beta^{2}\right)  \left(  \eta
_{1}\sin\theta+\eta_{2}\cos\theta\right)  -\operatorname{Re}\left(  \beta
^{2}\right)  \left(  \eta_{1}\cos\theta-\eta_{2}\sin\theta\right)  -\eta
_{3}\operatorname{Re}\left(  \alpha\beta e^{i\theta}\right)  \right)
X_{\theta}^{\ast}\label{dressing vector for g inverso}\\
&  ~~~~~-\frac{i}{2}\left(  \bar{\alpha}\beta\left(  \eta_{1}+i\eta
_{2}\right)  +\alpha\bar{\beta}\left(  \eta_{1}-i\eta_{2}\right)  \right)
X_{3}\nonumber
\end{align}
where $X_{\theta}=X_{1}\cos\theta+X_{2}\sin\theta$ and $X_{\theta}^{\ast
}=-X_{1}\sin\theta+X_{2}\cos\theta$. The orbit $\mathcal{O}_{\theta}=\left\{
x\left(  \cos\theta X_{1}+\sin\theta X_{2}\right)  +zX_{3}~/~x\in
\mathbb{R}_{>0},~z\in\mathbb{R}\right\}  $ can be characterized by means the
dual basis $\left\{  \mathbf{x}_{\theta},\mathbf{x}_{\theta}^{\ast}%
,\mathbf{x}_{3}\right\}  \subset\mathfrak{su}_{2}^{\ast}$, defining
$\mathbf{x}_{\theta}=\left(  \cos\theta\mathbf{x}_{1}+\sin\theta\mathbf{x}%
_{2}\right)  $ and $\mathbf{x}_{\theta}^{\ast}=\left(  -\sin\theta
\mathbf{x}_{1}+\cos\theta\mathbf{x}_{2}\right)  $, so that%
\[
\mathcal{O}_{\theta}=\left(  \mathbf{x}_{\theta}^{\ast}\right)  ^{\circ}%
\cap\mathbf{x}_{\theta}^{-1}\left(  \mathbb{R}_{>0}\right)
\]
In this way, we get for $\left(  \eqref{dressing vector for g inverso}\right)
$,
\begin{align*}
\left\langle \mathbf{x}_{\theta}^{\ast},g\left(  g^{-1}\right)  ^{\bar{\psi
}^{\ast}\left(  \eta\right)  }\right\rangle =  &  -\frac{i}{2}\left(
\beta^{2}-\bar{\beta}^{2}\right)  \left(  \eta_{1}\sin\theta+\eta_{2}%
\cos\theta\right) \\
&  -\frac{1}{2}\left(  \beta^{2}+\bar{\beta}^{2}\right)  \left(  \eta_{1}%
\cos\theta-\eta_{2}\sin\theta\right) \\
&  -\frac{1}{2}\eta_{3}\left(  \bar{\alpha}\bar{\beta}e^{-i\theta}+\alpha\beta
e^{i\theta}\right)
\end{align*}
The annihilator is obtained from the condition $\left\langle \mathbf{x}%
_{\theta}^{\ast},g\left(  g^{-1}\right)  ^{\bar{\psi}^{\ast}\left(
\eta\right)  }\right\rangle =0$, implying that $\operatorname{Re}\left(
\left(  \beta^{2}\eta_{+}+\eta_{3}\alpha\beta\right)  e^{i\theta}\right)  =0$,
where $\eta_{+}:=\eta_{1}+i\eta_{2}$. The remaining restriction is
$\left\langle \mathbf{x}_{\theta},g\left(  g^{-1}\right)  ^{\bar{\psi}^{\ast
}\left(  \eta\right)  }\right\rangle $ $>0$ that is equivalent to
$\operatorname{Im}\left(  \left(  \beta^{2}\eta_{+}+\eta_{3}\alpha
\beta\right)  e^{i\theta}\right)  <0$. Thus have shown the following statement.

\begin{description}
\item[Proposition:] $\tilde{\varphi}^{-1}\left(  \mathcal{O}_{\theta}\right)
$ \textit{is determined by the constraints }%
\begin{align}
\operatorname{Re}\left(  \left(  \beta^{2}\eta_{+}+\eta_{3}\alpha\beta\right)
e^{i\theta}\right)   &  =0\label{restr 1}\\
\operatorname{Im}\left(  \left(  \beta^{2}\eta_{+}+\eta_{3}\alpha\beta\right)
e^{i\theta}\right)   &  <0 \label{restr 2}%
\end{align}
\textit{on the components of }$\left(  g,\eta\right)  \in SU(2)\times
\mathfrak{su}_{2}^{\ast}$.
\end{description}

As explained above, the dualizable subspaces are the symplectic leaves in
$\tilde{\varphi}^{-1}\left(  \mathcal{O}_{\theta}\right)  $, which coincide
with the orbits of the action $\mathsf{\hat{d}}$, see $\left(
\ref{dress action}\right)  $. Despite the rather obscure description of
$\tilde{\varphi}^{-1}\left(  \mathcal{O}_{\theta}\right)  $, determining these
orbit looks simpler working out separately the factors in $TSU(2)$ and
$\mathfrak{su}_{2}^{\ast}$.

To start with, we work out the projection of $T\tilde{\varphi}^{-1}\left(
\mathcal{O}_{\theta}\right)  $ on the factor $TSU(2)$. Here, we make use of
the digression $\left(  \ref{orbits and bruhat}\right)  $ to conclude that
there are two kind of dressing orbits in $SU(2)$: the zero dimensional ones
determined by the points with $\beta=0$, and the two dimensional ones
determined by $\vartheta=\arg\beta$, for $\beta\neq0$, so they are two
dimensional spheres $S^{2}$. The zero dimensional case are trivial dualizable
subspaces, so we focus our attention on the last ones.

The infinitesimal generators $\left(  \ref{dress inf gen}\right)  $ involve
for each $g\in SU\left(  2\right)  $ the linear transformation associated to
$\mathsf{d}:\mathfrak{b}\longrightarrow TSU(2)\cong SU(2)\times\mathfrak{su}%
_{2}$ relating the basis $\left\{  H,E,iE\right\}  $ and $\left\{  X_{1}%
,X_{2},X_{3}\right\}  $, which has a non trivial \emph{kernel }spanned by the
vector
\[
\tilde{X}_{\circ}\left(  g\right)  =H-i\dfrac{1}{\left\vert \beta\right\vert
^{2}}\left(  \alpha\beta-\bar{\alpha}\bar{\beta}\right)  \left(  iE\right)
+\dfrac{1}{\left\vert \beta\right\vert ^{2}}\left(  \alpha\beta+\bar{\alpha
}\bar{\beta}\right)  E
\]
On the other side, since $\left(  \operatorname{Im}\mathsf{d}\right)  ^{\circ
}=\ker\mathsf{d}^{\top}$, where $\mathsf{d}^{\top}:SU(2)\times\mathfrak{su}%
_{2}^{\ast}\longrightarrow\mathfrak{b}^{\ast}$ is the \emph{transpose} of
$\mathsf{d}$, we make the identification $\operatorname{Im}\mathsf{d}=\left(
g,\hat{\pi}\right)  ^{\circ}$, with $\left(  g,\hat{\pi}\right)  \in
\ker\mathsf{d}^{\top}$ being the generator of $\ker\mathsf{d}^{\top}$ with
\begin{equation}
\hat{\pi}=-\dfrac{1}{\left\vert \beta\right\vert ^{2}}\operatorname{Re}\left(
\alpha\bar{\beta}\right)  \mathbf{x}_{1}-\dfrac{1}{\left\vert \beta\right\vert
^{2}}\operatorname{Im}\left(  \alpha\bar{\beta}\right)  \mathbf{x}%
_{2}+\mathbf{x}_{3} \label{dres orb annhilator}%
\end{equation}
written it in terms of the dual basis $\left\{  \mathbf{x}_{1},\mathbf{x}%
_{2},\mathbf{x}_{3}\right\}  \subset\mathfrak{su}_{2}^{\ast}$. So, $\hat{\pi}$
annihilates the vectors $\left(  \ref{dress inf gen}\right)  $ which spans the
tangent space of each dressing orbit.

The dressing action of $B$ on $SU(2)$ left invariant $\vartheta=\arg\beta$, so
it is a suitable parameter to characterize the dressing orbits in $SU(2)$, and
the Lie derivative along $\vartheta$ gives a normal vector to them. Thus, let
us consider this tangent vector
\begin{equation}
\left.  V_{0}\right\vert _{g}=\left(  g,-g^{-1}\dfrac{\partial g}%
{\partial\vartheta}\right)  =\left(  g,-\dfrac{1}{2}\left(  \alpha\bar{\beta
}+\bar{\alpha}\beta\right)  ~X_{1}-\frac{1}{2}i\left(  \bar{\alpha}%
\beta-\alpha\bar{\beta}\right)  ~X_{2}+\left\vert \beta\right\vert ^{2}%
~X_{3}\right)  \label{Lie deriv in theta minus}%
\end{equation}
One may easily verify that
\[
\left\langle \hat{\pi},-g^{-1}\dfrac{\partial g}{\partial\vartheta
}\right\rangle =1
\]

On the other side, the projection of $T\tilde{\varphi}^{-1}\left(
\mathcal{O}_{\theta}\right)  $ on the factor $\mathfrak{su}_{2}^{\ast}$ is
foliated by the orbits of the action
\[
\Pi_{\mathfrak{su}_{2}^{\ast}}\mathsf{\hat{d}}_{\tilde{b}}\left(
g,\eta\right)  =\left(  \psi^{\ast}\circ Ad_{\tilde{b}^{g}}\circ\bar{\psi
}^{\ast}\right)  \eta
\]
which left invariant the $\mathbf{x}_{3}$ component of $\eta$, turning
$\eta_{3}=\left\langle \eta,X_{3}\right\rangle $ in a good parameter for the
corresponding orbits.

Therefore, we introduce the projectors
\begin{align*}
\mathrm{P}_{0}  &  :TSU(2)\longrightarrow\Pi_{TSU(2)}T\tilde{\varphi}%
^{-1}\left(  \mathcal{O}_{\theta}\right)  ~/~\mathrm{P}_{0}=Id-V_{0}\left(
\hat{\pi}\circ\left(  L_{g^{-1}}\right)  _{*}\right) \\
& \\
\mathrm{P}_{3}  &  :T\mathfrak{su}_{2}^{\ast}\longrightarrow\Pi_{\mathfrak{su}%
_{2}^{\ast}}T\tilde{\varphi}^{-1}\left(  \mathcal{O}_{\theta}\right)
~/~\mathrm{P}_{3}=Id-\mathbf{x}_{3}X_{3}%
\end{align*}
so that $\mathrm{P}=P_{0}\times P_{3} $ is the projector onto the
$\mathsf{\hat{d}}$ orbits.

As mentioned before, the submanifold $\tilde{\varphi}^{-1}\left(
\mathcal{O}_{\theta}\right)  $ is a presymplectic one when endowed with the
restriction of the canonical form $\omega_{\circ}$ of $T^{\ast}SU(2)\cong
SU(2)\times\mathfrak{su}_{2}^{\ast}$, and its symplectic leaves are the orbits
of the action $\mathsf{\hat{d}}$ $\left(  \ref{dress action}\right)  $. We use
the above projectors to write down this restricted symplectic form on each
orbit starting from the relation
\[
\left\langle \omega_{\circ}^{R},\left(  v,\rho\right)  \otimes\left(
w,\lambda\right)  \right\rangle _{\left(  g,\eta\right)  \in\tilde{\varphi
}^{-1}\left(  \mathcal{O}_{\theta}\right)  }=\left\langle \omega_{\circ
},\mathrm{P}\left(  v,\rho\right)  \otimes\mathrm{P}\left(  w,\lambda\right)
\right\rangle _{\left(  g,\eta\right)  \in\tilde{\varphi}^{-1}\left(
\mathcal{O}_{\theta}\right)  }%
\]
to get the expression
\begin{equation}
\omega_{\circ}^{R}=\omega_{\circ}-L_{g^{-1}\ast}\hat{\pi}\wedge\left(
g^{-1}V_{0}-\left\vert \beta\right\vert ^{2}X_{3}+g^{-1}ad_{g^{-1}V_{0}}%
^{\ast}\eta\right)  +X_{3}\wedge g^{-1}\mathbf{x}_{3}
\label{red sympl form in T*SU(2)}%
\end{equation}
from where we obtain the Dirac bracket%
\begin{align*}
\left\{  \mathcal{F},\mathcal{G}\right\}  _{\tilde{\varphi}^{-1}\left(
\mathcal{O}_{\theta}\right)  }^{D}\left(  g,\eta\right)  =  &  \left\{
\mathcal{F},\mathcal{G}\right\}  \left(  g,\eta\right)  -\left\langle
g\mathbf{d}\mathcal{F},\left\langle \hat{\pi},\delta\mathcal{G}\right\rangle
X_{3}\right\rangle +\left\langle g\mathbf{d}\mathcal{G},\left\langle \hat{\pi
},\delta\mathcal{F}\right\rangle X_{3}\right\rangle \\
&  +\left\langle \eta-\dfrac{\left\langle \eta,g^{-1}V_{0}\right\rangle
}{\left\vert \beta\right\vert ^{2}}\mathbf{x}_{3},\left[  \delta
\mathcal{F},\delta\mathcal{G}\right]  \right\rangle
\end{align*}
and the hamiltonian vector field
\begin{equation}
V_{\mathcal{G}}=V_{\mathcal{G}}^{\circ}-\left(  \left\langle \hat{\pi}%
,\delta\mathcal{G}\right\rangle X_{3},ad_{\delta\mathcal{G}}^{\ast}\left(
\eta-\dfrac{\left\langle \eta,g^{-1}V_{0}\right\rangle }{\left\vert
\beta\right\vert ^{2}}\mathbf{x}_{3}\right)  -\left\langle g\mathbf{d}%
\mathcal{G},X_{3}\right\rangle \hat{\pi}\right)  \label{restr ham vec field}%
\end{equation}
which is tangent to the $\mathsf{\hat{d}}$-orbits in $\tilde{\varphi}%
^{-1}\left(  \mathcal{O}_{\theta}\right)  $, as expected.

\section{AKS integrability scheme and dynamics on factors of $SL\left(
2,\mathbb{C}\right)  $}

Let us work out the dynamical setting on the coadjoint orbit in $\mathfrak{sl}%
_{2}^{\ast}\mathbb{C}$, which we shall identify $\mathfrak{sl}_{2}\mathbb{C}$
through the invariant non degenerate bilinear form $(X,Y)_{\mathfrak{sl}_{2}}%
$, $\left(  \ref{prod-escal 0}\right)  $. Following the AKS scheme, we choose
an $\mathrm{Ad}$-invariant function on the a coadjoint orbit of $SL\left(
2,\mathbb{C}\right)  $, so that its restriction to the coadjoint orbit of $B$
in $\mathfrak{b}^{\ast}\hookrightarrow\mathfrak{sl}_{2}^{\ast}\mathbb{C}$
gives a nontrivial dynamics embracing also the dynamics on the cotangent
bundles $T^{\ast}SU(2)$ and $T^{\ast}B$ $\left(  \cite{RSTS 1},\cite{RSTS
2}\right)  $. In particular, the $\mathrm{Ad}$-invariant function
$f:\mathfrak{sl}_{2}\mathbb{C\longrightarrow R}$
\[
f\left(  \mathbf{X}\right)  :=-\frac{1}{16}\mathrm{\operatorname{Re}}%
\kappa\left(  \mathbf{X},\mathbf{X}\right)
\]
is related to the Hamiltonian function of the Toda model, as we shall see
below. Its Legendre transform, $\mathcal{L}_{f}:\mathfrak{sl}_{2}%
\mathbb{C}\longrightarrow\mathfrak{sl}_{2}^{\ast}\mathbb{C}$, is%
\[
\left\langle \mathcal{L}_{f}\left(  \mathbf{X}\right)  ,\mathbf{Y}%
\right\rangle \equiv\left.  \frac{\mathrm{d}}{\mathrm{d}t}f\left(
\mathbf{X}+t\mathbf{Y}\right)  \right\vert _{t=0}=-\frac{1}{8}%
\mathrm{\operatorname{Re}}\kappa\left(  \mathbf{X},\mathbf{Y}\right)
\]
valid for all $\mathbf{Y}\in\mathfrak{sl}\left(  2,\mathbb{C}\right)  $. As an
element of $\mathfrak{sl}\left(  2,\mathbb{C}\right)  $ through the invariant
product $\left(  \cdot,\cdot\right)  _{\mathfrak{sl}_{2}}$, it is given by
$\left(  \mathcal{\tilde{L}}_{f}\left(  \mathbf{X}\right)  ,\mathbf{Y}\right)
_{\mathfrak{sl}_{2}}=-\frac{1}{4}\mathrm{\operatorname{Im}}\,\kappa
(\mathcal{\tilde{L}}_{f}\left(  \mathbf{X}\right)  ,\mathbf{Y})$ and the
$\mathbb{C}$-linearity of $\kappa$ enables to make the identification
$\operatorname{Re}\left(  \kappa\left(  \mathbf{X},\mathbf{Y}\right)  \right)
=\operatorname{Im}\left(  \kappa\left(  i\mathbf{X},\mathbf{Y}\right)
\right)  $. Hence, from the definition of the bilinear product $\left(
\cdot,\cdot\right)  _{\mathfrak{sl}_{2}}$, we get%
\[
\operatorname{Im}\kappa\left(  \mathcal{\tilde{L}}_{f}\left(  \mathbf{X}%
\right)  -\frac{i}{2}\mathbf{X},\mathbf{Y}\right)  =0,\qquad\forall
\mathbf{Y}\in\mathfrak{sl}\left(  2,\mathbb{C}\right)
\]
and, because of the non degeneracy of $\left(  \cdot,\cdot\right)
_{\mathfrak{sl}_{2}}$, we conclude that
\[
\mathcal{\tilde{L}}_{f}\left(  \mathbf{X}\right)  =\frac{i}{2}\mathbf{X}%
\]

Applying the AKS scheme to $T^{\ast}SU(2)$ and $T^{\ast}B$, regarded as
$B$-hamiltonian spaces, requires that $\psi^{\ast}\left(  \Pi_{\mathfrak{b}%
}\mathbf{X}\right)  $ be a character of $\mathfrak{su}_{2}$. The only chance
is $\Pi_{\mathfrak{b}}\mathbf{X}=0$, so $\mathbf{X}\equiv X\in\mathfrak{su}%
_{2}$ and the hamiltonian vector field associated with $f$ has integral curves
given by the \emph{fundamental flux} $t\mapsto\exp t\mathcal{\tilde{L}}%
_{f}\left(  X\right)  $: if $X=a_{1}X_{1}+a_{2}X_{2}+a_{3}X_{3}$,
\[
\exp t\mathcal{\tilde{L}}_{f}\left(  X\right)  =\cosh\left(  t\frac{\left\vert
\left\vert X\right\vert \right\vert }{2}\right)  +iX\sinh\left(
t\frac{\left\vert \left\vert X\right\vert \right\vert }{2}\right)
\]
Here $\left\vert \left\vert X\right\vert \right\vert =\sqrt{\mathrm{det}X}$,
and from now on we consider $\mathrm{det}X=1$, equivalent to $a_{1}^{2}%
+a_{2}^{2}+a_{3}^{2}=1$. Thus, the solutions to collective $B$-hamiltonian
systems are orbits of a curve in $B$ obtained from the factorization of the
curve $\exp t\mathcal{\tilde{L}}_{f}\left(  X\right)  $ on $SU(2)\times B$
$\left(  \cite{RSTS 1},\cite{RSTS 2}\right)  $. In fact, the curve in
$SL(2,\mathbb{C)}$%
\begin{equation}
\exp t\mathcal{\tilde{L}}_{f}\left(  X\right)  =\left[
\begin{array}
[c]{cc}%
\cosh\left(  t/2\right)  -a_{3}\sinh\left(  t/2\right)  & -\left(
a_{1}-ia_{2}\right)  \sinh\left(  t/2\right) \\
-\left(  a_{1}+ia_{2}\right)  \sinh\left(  t/2\right)  & \cosh\left(
t/2\right)  +a_{3}\sinh\left(  t/2\right)
\end{array}
\right]  \label{ezp curve}%
\end{equation}
can be factorized out as
\[
\exp t\mathcal{\tilde{L}}_{f}\left(  X\right)  =g\left(  t\right)  \tilde
{b}\left(  t\right)
\]
with $g\left(  t\right)  \subset SU(2)$ and $\tilde{b}\left(  t\right)
\subset B$ given by%
\begin{align}
g\left(  t\right)   &  =\left[
\begin{array}
[c]{ccc}%
\dfrac{\cosh{\left(  t/2\right)  }-a_{3}\sinh{\left(  t/2\right)  }}%
{\sqrt{\cosh{t}-a_{3}\sinh{t}}} &  & \dfrac{\left(  a_{1}-ia_{2}\right)
\sinh{\left(  t/2\right)  }}{\sqrt{\cosh{t}-a_{3}\sinh{t}}}\\
&  & \\
-\dfrac{\left(  a_{1}+ia_{2}\right)  \sinh{\left(  t/2\right)  }}{\sqrt
{\cosh{t}-a_{3}\sinh{t}}} &  & \dfrac{\cosh{\left(  t/2\right)  }-a_{3}%
\sinh{\left(  t/2\right)  }}{\sqrt{\cosh{t}-a_{3}\sinh{t}}}%
\end{array}
\right] \label{SU(2) factor}\\
& \nonumber\\
\tilde{b}\left(  t\right)   &  =\left[
\begin{array}
[c]{ccc}%
\sqrt{\cosh{t}-a_{3}\sinh{t}} &  & \dfrac{-\left(  a_{1}-ia_{2}\right)
\sinh{t}}{\sqrt{\cosh{t}-a_{3}\sinh{t}}}\\
&  & \\
0 &  & \left(  \sqrt{\cosh{t}-a_{3}\sinh{t}}\right)  ^{-1}%
\end{array}
\right]  \label{B factor}%
\end{align}
Hence, the solution curves of suitable $B$-hamiltonian systems are given by
the orbits of $\tilde{b}\left(  t\right)  $ in each space, as we shall
describe below.

\section{Dynamics on $B$-spaces}

T-duality relates dynamical systems on the three hamiltonian $B$-spaces,
namely $T^{\ast}B$, $T^{\ast}SU\left(  2\right)  $ and $\mathbb{R}^{2}$. The
restriction of the $\mathrm{Ad}^{\ast}$-invariant function $f$ to $\left(
\mathfrak{su}_{2}\right)  ^{\circ}\cong\mathfrak{su}_{2}$ induces, on each of
these spaces, collective systems whose solution can be found through the AKS
method. Accordingly with it, we just need to know the form of the action of
the Lie group $B$ on the spaces under consideration to find out the solution
to the equation of motion. .

Hamiltonian systems modelled on the cotangent bundles of a Lie group $G$ are
characterized, in body coordinates, by the canonical symplectic structure
which, besides the Hamilton function, defines equations through the
hamiltonian vector field $\left.  V_{\mathcal{H}}\right\vert _{\left(
g,\eta\right)  }=\left(  g\delta\mathcal{H},ad_{\delta\mathcal{H}}^{\ast}%
\eta-g\mathbf{d}\mathcal{H}\right)  $, for the function $\mathcal{H}\in
C^{\infty}\left(  G\times\mathfrak{g}^{\ast}\right)  $, where $d\mathcal{H}%
=\left(  \mathbf{d}\mathcal{H},\delta\mathcal{H}\right)  $, with
$\mathbf{d}\mathcal{H}\in T_{g}^{\ast}G$ and $\delta\mathcal{H}\in T_{\eta
}^{\ast}\mathfrak{g}^{\ast}$, so
\[%
\begin{array}
[c]{l}%
g^{-1}\dot{g}=\delta\mathcal{H}\\
\\
\dot{\eta}=ad_{\delta\mathcal{H}}^{\ast}\eta-g\mathbf{d}\mathcal{H}%
\end{array}
\]
As explained, we consider a function $\mathsf{h}:\mathfrak{su}_{2}%
\longrightarrow\mathbb{R}$ and, in each case, the hamiltonian functions of the
respective dynamical systems are obtained by composing them with the
corresponding momentum maps.

\subsection{Dynamics on $\mathbb{R}^{2}$}

The action of $B$ on $\mathbb{R}^{2}$, eq. $\left(  \ref{rho 0}\right)  $, is
\[
\rho\left(  \tilde{b},(q,p)\right)  =\left(  q-\frac{1}{\mu}\ln a,p-2\mu
\dfrac{\varepsilon}{a}\exp(2\mu q)\left(  b\cos\theta-c\sin\theta\right)
\right)
\]
where
\[
\tilde{b}=\left[
\begin{array}
[c]{cc}%
a & b+ic\\
0 & a^{-1}%
\end{array}
\right]
\]
with associated momentum map $\tilde{\sigma}_{\theta}$
\[
\tilde{\sigma}_{\theta}(q,p)=\dfrac{1}{2\mu}pX_{3}-\varepsilon\exp(2\mu
q)(\cos\theta X_{1}+\sin\theta X_{2})
\]
The collective hamiltonian here is
\[
\mathcal{H}_{\mathbb{R}^{2}}\left(  q,p\right)  =-\frac{1}{16}%
\operatorname{Re}\kappa\left(  \tilde{\sigma}_{\theta}(q,p),\tilde{\sigma
}_{\theta}(q,p)\right)  =\frac{1}{2}\left(  \frac{1}{4\mu^{2}}p^{2}%
+2\varepsilon^{2}\exp{\left(  4\mu q\right)  }\right)
\]
By choosing the point $X_{\circ}:=n_{\circ}\left(  \cos{\theta}X_{1}%
+\sin{\theta}X_{2}\right)  +m_{\circ}X_{3}$, $n_{\circ}^{2}+m_{\circ}^{2}=1$,
in $\mathcal{O}_{\theta}\subset\mathfrak{su}_{2}$, the solution curve through
the initial point $\left(  q_{\circ},p_{\circ}\right)  $ with $\tilde{\sigma
}_{\theta}(q_{\circ},p_{\circ})=X_{\circ}$ is given by the action of the curve
$\tilde{b}\left(  t\right)  $, obtained in eq. $\left(  \ref{B factor}\right)
$,
\[
t\mapsto\rho\left(  \tilde{b}\left(  t\right)  ,(q_{\circ},p_{\circ})\right)
=\left(  q_{\circ}-\frac{1}{\mu}\ln a\left(  t\right)  ,p_{\circ}-\dfrac
{2\mu\varepsilon}{a\left(  t\right)  }e^{2\mu q_{\circ}}\left(  b\left(
t\right)  \cos\theta-c\left(  t\right)  \sin\theta\right)  \right)
\]
where%
\begin{align*}
a\left(  t\right)   &  =\sqrt{\cosh{t}-a_{3}\sinh{t}}\\
b\left(  t\right)   &  =\dfrac{-a_{1}\sinh{t}}{\sqrt{\cosh{t}-a_{3}\sinh{t}}%
}\\
c\left(  t\right)   &  =\dfrac{a_{2}\sinh{t}}{\sqrt{\cosh{t}-a_{3}\sinh{t}}}%
\end{align*}
and%
\begin{align*}
a_{1}  &  =-\varepsilon\exp(2\mu q)\cos\theta\\
a_{2}  &  =-\varepsilon\exp(2\mu q)\sin\theta\\
a_{3}  &  =\dfrac{1}{2\mu}p
\end{align*}
so%
\begin{align*}
\rho\left(  \tilde{b}\left(  t\right)  ,(q_{\circ},p_{\circ})\right)  =  &
\left(  q_{\circ}-\frac{1}{2\mu}\ln\left(  \cosh{t}-\dfrac{1}{2\mu}p\sinh
{t}\right)  ,\right. \\
&  \left.  p_{\circ}-2\mu\dfrac{\varepsilon^{2}}{\left(  \cosh{t}-\dfrac
{1}{2\mu}p\sinh{t}\right)  }\exp(2\mu\left(  q_{\circ}+q\right)  )\sinh
{t}\right)
\end{align*}

In order to have $\tilde{\sigma}_{\theta}(q_{\circ},p_{\circ})=X_{\circ}$, the
following relations must hold
\[
\left.
\begin{array}
[c]{l}%
m_{\circ}=\frac{1}{2\mu}p_{\circ}\\
n_{\circ}=-\varepsilon\exp(2\mu q_{\circ})
\end{array}
\right\}  \Longrightarrow\left(  \frac{1}{2\mu}p_{\circ}\right)
^{2}+\varepsilon^{2}\exp(4\mu q_{\circ})=1
\]
Thus, the curve $\rho\left(  \tilde{b}\left(  t\right)  ,(q_{\circ},p_{\circ
})\right)  $ becomes in%

\[
\rho\left(  \tilde{b}\left(  t\right)  ,(q_{\circ},p_{\circ})\right)  =\left(
q_{\circ}-\frac{1}{2\mu}\ln\left(  \cosh{t}-\frac{1}{2\mu}p_{\circ}\sinh
{t}\right)  ,-2\mu\dfrac{\sinh{t}-\frac{1}{2\mu}p_{\circ}\cosh{t}}{\cosh
{t}-\frac{1}{2\mu}p_{\circ}\sinh{t}}\right)
\]

For the particular values $\mu=\dfrac{1}{2}$ and $\varepsilon=\pm\sqrt{2}$ in
$\mathcal{H}_{\mathbb{R}^{2}}\left(  q,p\right)  $, we obtain the \emph{Toda
hamiltonian}%

\begin{equation}
\mathcal{H}_{Toda}\left(  q,p\right)  =\frac{1}{2}p^{2}+\exp(2q)
\label{Toda Ham}%
\end{equation}
whose Hamilton equation
\[
\left\{
\begin{array}
[c]{l}%
\dot{q}=p\\
\\
\dot{p}=-2\exp{(2q)}%
\end{array}
\right.
\]
are solved by the curves
\[%
\begin{array}
[c]{ccc}%
q\left(  t\right)  =-\ln\left(  \cosh{t}-p_{\circ}\sinh{t}\right)  & , &
p\left(  t\right)  =-\dfrac{\sinh{t}-p_{\circ}\cosh{t}}{\cosh{t}-p_{\circ
}\sinh{t}}%
\end{array}
\]
\bigskip

\begin{description}
\item[A little note about parameters.] \textit{In the previous setting it was
possible to solve the Toda eqs of motion in case in which the energy of the
system is equal to }$1$\textit{, however it is possible to choose the
parameters in order to solve the system at any other (positive of course)
energy.}
\end{description}

The lagrangian corresponding to the Toda hamiltonian $\left(  \ref{Toda Ham}%
\right)  $ is
\begin{equation}
\mathcal{L}_{Toda}\left(  q,\dot{q}\right)  =\frac{1}{2}\dot{q}^{2}-\exp(2q)
\label{Toda Lag}%
\end{equation}

\subsection{Dynamics on $T^{*}B$}

In terms of the momentum map $\tilde{\mu}\left(  \tilde{b},\tilde{\eta
}\right)  =\left(  \bar{\psi}\circ\mathrm{Ad}_{\tilde{b}^{-1}}^{\ast}\circ
\psi\right)  \psi\left(  \tilde{\eta}\right)  $, given in eq. $\left(
\ref{m-map in T*B}\right)  $, the collective hamiltonian on $T^{\ast}B$ is
then%
\begin{align*}
\mathcal{H}_{B}\left(  \tilde{b},\tilde{\eta}\right)   &  =-\frac{1}{16}%
\kappa\left(  \tilde{\mu}\left(  \tilde{b},X\right)  ,\tilde{\mu}\left(
\tilde{b},X\right)  \right) \\
&  =\frac{1}{2a^{4}}\left(  \tilde{\eta}_{\mathbf{e}}^{2}+\tilde{\eta
}_{\mathbf{\tilde{e}}}^{2}\right)  +\frac{1}{2}\left(  \frac{1}{2}\tilde{\eta
}_{\mathbf{h}}+\dfrac{b\tilde{\eta}_{\mathbf{e}}+c\tilde{\eta}_{\mathbf{\tilde
{e}}}}{a}\right)  ^{2}%
\end{align*}
The evolution curve passing through the initial point $\left(  \tilde
{b}_{\circ},X_{\circ}\right)  $ with $X_{\circ}=\left(  n_{\circ}\cos
\theta~X_{1}+n_{\circ}\sin\theta~X_{2}\right)  +m_{\circ}X_{3}$ is
\[
t\mapsto\tilde{\lambda}\left(  \tilde{b}\left(  t\right)  ,\left(  \tilde
{b}_{\circ},\tilde{\eta}_{\circ}\right)  \right)  =\left(  \left[
\begin{array}
[c]{cc}%
a\left(  t\right)  & z\left(  t\right) \\
0 & \left(  a\left(  t\right)  \right)  ^{-1}%
\end{array}
\right]  \cdot\left[
\begin{array}
[c]{cc}%
a_{\circ} & z_{\circ}\\
0 & a_{\circ}%
\end{array}
\right]  ,X_{\circ}\right)
\]
where the curve $\tilde{b}\left(  t\right)  $ is that of eq. $\left(
\ref{B factor}\right)  $. Explicitly, it means the curve
\begin{align*}
&  \tilde{\lambda}\left(  \tilde{b}\left(  t\right)  ,\left(  \tilde{b}%
_{\circ},\tilde{\eta}_{\circ}\right)  \right) \\
&  =\left(  \left[
\begin{array}
[c]{cc}%
a_{\circ}\sqrt{\cosh{t}-m_{\circ}\sinh{t}} & \dfrac{a_{\circ}z_{\circ}\cosh
{t}-\left(  m_{\circ}a_{\circ}z_{\circ}+n_{\circ}e^{-i\theta}\right)  \sinh
{t}}{a_{\circ}\sqrt{\cosh{t}-m_{\circ}\sinh{t}}}\\
0 & \dfrac{1}{a_{\circ}\sqrt{\cosh{t}-m_{\circ}\sinh{t}}}%
\end{array}
\right]  ,X_{\circ}\right)
\end{align*}

The lagrangian version of this model can be retrieved from the first Hamilton
equation
\[
\tilde{b}^{-1}\overset{\cdot}{\tilde{b}}=\delta\mathcal{H}_{B}%
\]
that in explicit form is
\begin{align*}
a\dot{b}-\dot{a}b &  =\tilde{\eta}_{\mathbf{e}}a^{-2}\\
& \\
a\dot{c}-\dot{a}c &  =\tilde{\eta}_{\mathbf{\tilde{e}}}a^{-2}\\
& \\
a^{-1}\dot{a} &  =\frac{1}{2}\left(  \tilde{\eta}_{\mathbf{e}}ba^{-1}%
+\tilde{\eta}_{\mathbf{\tilde{e}}}ca^{-1}+\frac{1}{2}\tilde{\eta}_{\mathbf{h}%
}\right)
\end{align*}
Hence, the Lagrangian is obtained as%
\[
\tilde{L}_{B}\left(  \tilde{b},\overset{\cdot}{\tilde{b}}\right)
=\left\langle \tilde{\eta},\tilde{b}^{-1}\overset{\cdot}{\tilde{b}%
}\right\rangle -\mathcal{H}_{B}\left(  \tilde{b},\tilde{\eta}\right)
\]
that after some calculation gives%
\begin{equation}
\tilde{L}_{B}\left(  \tilde{b},\overset{\cdot}{\tilde{b}}\right)  =\frac{1}%
{2}\left(  b\dot{a}-a\dot{b}\right)  ^{2}+\frac{1}{2}\left(  c\dot{a}-a\dot
{c}\right)  ^{2}+2\left(  a^{-1}\dot{a}\right)  ^{2}%
\label{full lagrangian for B}%
\end{equation}
This Lagrangian reduce on $\tilde{\mu}^{-1}\mathcal{O}_{\theta}$ to
\[
\tilde{L}_{B}^{red}\left(  \tilde{b},\overset{\cdot}{\tilde{b}}\right)
=\frac{1}{2}\left(  t\dot{a}-a\dot{t}\right)  ^{2}+2\left(  \dfrac{\dot{a}}%
{a}\right)  ^{2}%
\]

Since%
\[
\overset{\cdot}{\tilde{b}}\tilde{b}^{-1}=\left(  a\dot{b}-\dot{a}b\right)
E+\left(  a\dot{c}-\dot{a}c\right)  \left(  iE\right)  +a^{-1}\dot{a}H
\]
and introducing the linear map $\mathbb{K}:=\hat{\kappa}_{\mathfrak{su}_{2}%
}^{-1}\circ\psi^{\ast}:\mathfrak{b}\longrightarrow\mathfrak{su}_{2}$, given by%
\[
\mathbb{K}E=\frac{1}{8}X_{1},\qquad\mathbb{K}\left(  iE\right)  =-\frac{1}%
{8}X_{2},\qquad\mathbb{K}H=\frac{1}{4}X_{3}%
\]
we may write the Lagrangian function $\left(  \ref{full lagrangian for B}%
\right)  $ using the symmetric bilinear form $\left(  \ref{prod-escal 0}%
\right)  $ on $\mathfrak{sl}_{2}$ as
\begin{equation}
\tilde{L}_{B}\left(  \tilde{b},\overset{\cdot}{\tilde{b}}\right)
=-4(\mathbb{K}\overset{\cdot}{\tilde{b}}\tilde{b}^{-1},\overset{\cdot}%
{\tilde{b}}\tilde{b}^{-1})_{\mathfrak{sl}_{2}}\label{full lagrangian for B II}%
\end{equation}
that resembles a generalized top lagrangian on $B$.

\subsection{Dynamics on $T^{\ast}SU\left(  2\right)  $}

We consider the hamiltonian function
\begin{equation}
\mathcal{H}_{SU\left(  2\right)  }(g,\eta)=-\dfrac{1}{16}\kappa(\tilde
{\varphi}(g,\eta),\tilde{\varphi}(g,\eta))
\label{hamiltonian for the dressing model}%
\end{equation}
In order to get the Hamilton equation of motion, we need to calculate the
differential $\left(  \mathbf{d}\mathcal{H}_{SU\left(  2\right)  }%
,\delta\mathcal{H}_{SU\left(  2\right)  }\right)  $. In doing so, we use the
expression for the differential of the momentum map $\tilde{\varphi}$
\[
\tilde{\varphi}_{\ast}\left(  gX,\xi\right)  =-\left(  id\otimes Ad_{g^{-1}%
}^{\ast}\xi\right)  \pi^{R}\left(  g\right)  +\left(  id\otimes Ad_{g^{-1}%
}^{\ast}\eta\right)  \left(  id\otimes ad_{X}\right)  \pi^{R}\left(  g\right)
-\left(  Ad_{g}\otimes\eta\right)  \delta_{\mathfrak{su}_{2}}(X)
\]
where $\delta_{\mathfrak{su}_{2}}:\mathfrak{su}_{2}\longrightarrow
\mathfrak{su}_{2}\otimes\mathfrak{su}_{2}$ is the coboundary coalgebra
structure of $\mathfrak{su}_{2}$ and $\pi^{R}\left(  g\right)  =\left(
R_{g^{-1}\ast}\otimes R_{g^{-1}\ast}\right)  \pi_{SU\left(  2\right)  }\left(
g\right)  $. Thus, the differential of the Hamilton function are%
\begin{align*}
g\mathbf{d}\mathcal{H}_{SU\left(  2\right)  }  &  =\dfrac{1}{8}\left[
\hat{\kappa}_{\mathfrak{su}_{2}}\tilde{\varphi}(g,\eta),Ad_{g^{-1}}^{\ast
}\left(  \bar{\psi}^{\ast}\left(  \eta\right)  \right)  \right]
_{\mathfrak{su}_{2}^{\ast}}\\
& \\
\delta\mathcal{H}_{SU\left(  2\right)  }  &  =\dfrac{1}{8}Ad_{g^{-1}}\left(
\hat{\kappa}_{\mathfrak{su}_{2}}\left(  \tilde{\varphi}(g,\eta)\right)
\otimes id\right)  \pi^{R}\left(  g\right)
\end{align*}
Observe that%

\begin{equation}
\delta\mathcal{H}_{SU\left(  2\right)  }=\dfrac{1}{8}Ad_{g^{-1}}\left(
\hat{\kappa}_{\mathfrak{su}_{2}}\left(  \tilde{\varphi}(g,\eta)\right)
\otimes id\right)  \pi^{R}\left(  g\right)  =g^{-1}g^{\bar{\psi}^{\ast}\left(
\hat{\kappa}_{\mathfrak{su}_{2}}\left(  \tilde{\varphi}(g,\eta)\right)
\right)  }\nonumber
\end{equation}
so $\left\langle \hat{\pi},\delta\mathcal{H}_{SU\left(  2\right)
}\right\rangle =0$.

Now, using the expression for the hamiltonian vector field on the
$\mathsf{\hat{d}}$-orbits $\left(  \ref{restr ham vec field}\right)  $%
\[
V_{\mathcal{G}}=V_{\mathcal{G}}^{\circ}-\left(  \left\langle \hat{\pi}%
,\delta\mathcal{G}\right\rangle X_{3},ad_{\delta\mathcal{G}}^{\ast}\left(
\eta-\dfrac{\left\langle \eta,g^{-1}V_{0}\right\rangle }{\left\vert
\beta\right\vert ^{2}}\mathbf{x}_{3}\right)  -\left\langle g\mathbf{d}%
\mathcal{G},X_{3}\right\rangle \hat{\pi}\right)
\]
where $V_{\mathcal{G}}^{\circ}=\left(  g\delta\mathcal{G},ad_{\delta
\mathcal{G}}^{\ast}\lambda-g\mathbf{d}\mathcal{G}\right)  $ is the hamiltonian
vector field associated with the canonical Poisson bracket in $TSU\left(
2\right)  $. We get
\[
V_{\mathcal{H}_{SU\left(  2\right)  }}=\left(  g\delta\mathcal{H}_{SU\left(
2\right)  },ad_{\delta\mathcal{H}_{SU\left(  2\right)  }}^{\ast}\left(
\dfrac{\left\langle \eta,g^{-1}V_{0}\right\rangle }{\left\vert \beta
\right\vert ^{2}}\mathbf{x}_{3}\right)  -g\mathbf{d}\mathcal{H}_{SU\left(
2\right)  }\right)
\]
which, obviously, satisfy $\mathrm{P}V_{\mathcal{H}_{SU\left(  2\right)  }%
}=V_{\mathcal{H}_{SU\left(  2\right)  }}$, for the projector $\mathrm{P}%
=\left(  Id-V_{0}\hat{\pi}\right)  \oplus\left(  Id-\mathbf{x}_{3}%
X_{3}\right)  $ on the $\mathsf{\hat{d}}$-orbits. So the reduced equation of
motion are
\[
\left\{
\begin{array}
[c]{l}%
g^{-1}\dot{g}=g\delta\mathcal{H}_{SU\left(  2\right)  }\\
\\
\dot{\eta}=\dfrac{1}{\left\vert \beta\right\vert ^{2}}\left\langle \eta
,g^{-1}V_{0}\right\rangle ad_{\delta\mathcal{H}_{SU\left(  2\right)  }}^{\ast
}\mathbf{x}_{3}-g\mathbf{d}\mathcal{H}_{SU\left(  2\right)  }%
\end{array}
\right.
\]
Observe that the equation for $g$, including the explicit form of
$\delta\mathcal{H}_{SU\left(  2\right)  }$ given above in terms of $\pi
^{R}\left(  g\right)  $, resembles the equation of motion of a finite
dimensional Poisson sigma model.

The solution curve is generated by the action of the curve $\tilde{b}\left(
t\right)  $, eq. $\left(  \ref{B factor}\right)  $, through the dressing
action $\mathsf{\hat{d}}:B\times SU\left(  2\right)  \times\mathfrak{su}%
_{2}^{\ast}\longrightarrow SU\left(  2\right)  \times\mathfrak{su}_{2}^{\ast}%
$. Having in mind the explicit form of $\tilde{\varphi}(g,\eta)$, we consider
the initial point $(g_{\circ},\eta_{\circ})$ such that $\tilde{\varphi}\left(
g_{\circ},\eta_{\circ}\right)  =X_{\circ}:=n_{\circ}\left(  \cos\theta
X_{1}+\sin\theta X_{2}\right)  +m_{\circ}X_{3}$, $n_{\circ}^{2}+m_{\circ}%
^{2}=1$, in $\mathcal{O}_{\theta}\subset\mathfrak{su}_{2}$. A suitable
election for the initial condition is%
\[
g_{\circ}\left(  \psi_{\circ},\phi_{\circ}\right)  =\left(
\begin{array}
[c]{cc}%
0 & -e^{i\left(  \phi_{\circ}-\psi_{\circ}\right)  }\\
e^{-i\left(  \phi_{\circ}-\psi_{\circ}\right)  } & 0
\end{array}
\right)  ~~,~~\eta_{\circ}=\eta_{3\circ}\left(
\begin{array}
[c]{c}%
-\sin2\psi_{\circ}\\
\cos2\psi_{\circ}\\
1
\end{array}
\right)
\]
with
\[
\eta_{3\circ}^{2}\cos^{2}\left(  \theta+2\phi_{\circ}\right)  =1
\]
that gives a solution curves in each $\mathsf{\hat{d}}$-orbit
\[
\left(  g\left(  t\right)  ,\eta\left(  t\right)  \right)  =\left(  g_{\circ
}^{\tilde{b}\left(  t\right)  },\left(  \psi^{\ast}\circ Ad_{\left(  \tilde
{b}\left(  t\right)  \right)  ^{g_{\circ}}}\circ\bar{\psi}^{\ast}\right)
\eta_{\circ}\right)
\]
with%
\begin{align*}
g_{\circ}^{\tilde{b}\left(  t\right)  }=  &  \dfrac{1}{\sqrt{\lambda^{2}%
\sinh^{2}{t}+1}}\left(
\begin{array}
[c]{cc}%
-\lambda e^{-i\left(  \phi_{\circ}-\psi_{\circ}+\sigma\right)  }\sinh{t} &
-e^{i\left(  \phi_{\circ}-\psi_{\circ}\right)  }\\
e^{-i\left(  \phi_{\circ}-\psi_{\circ}\right)  } & -\lambda e^{i\left(
\phi_{\circ}-\psi_{\circ}+\sigma\right)  }\sinh{t}%
\end{array}
\right) \\
& \\
\eta\left(  t\right)  =  &  -\eta_{3}\left(  \sin\left(  2\psi\right)  \left(
\cosh{t}+a_{3}\sinh{t}\right)  +\lambda\cos\left(  \sigma-2\left(  \phi
_{\circ}-\psi_{\circ}\right)  \right)  \sinh{t}\right)  \mathbf{x}_{1}\\
&  +\eta_{3}\left(  \cos\left(  2\psi\right)  \left(  \cosh{t}+a_{3}\sinh
{t}\right)  +\lambda\sin\left(  \sigma-2\left(  \phi_{\circ}-\psi_{\circ
}\right)  \right)  \sinh{t}\right)  \mathbf{x}_{2}+\eta_{3}\mathbf{x}_{3}%
\end{align*}
Here, we wrote the paramenters $a_{1},a_{2}$ in the curve $\tilde{b}\left(
t\right)  $, eq. $\left(  \ref{B factor}\right)  $, as: $a_{1}=\lambda
\cos\sigma$ and $a_{2}=\lambda\sin\sigma$.

Let us to obtain the lagrangian version of this system. The Legendre
transformation in this case is singular, it can be partially retrieved from
the first Hamilton equation written as%
\[
g^{-1}\dot{g}=-g^{-1}g^{\bar{\psi}^{\ast}\left(  \hat{\kappa}_{\mathfrak{su}%
_{2}}\left(  \tilde{\varphi}(g,\eta)\right)  \right)  }%
\]
Explicitly, this equation are%
\begin{align*}
\bar{\alpha}\dot{\beta}-\beta\overset{\cdot}{\bar{\alpha}} &  =-\beta
^{2}\left(  \tilde{\varphi}_{1}(g,\eta)+i\tilde{\varphi}_{2}(g,\eta)\right)
+\bar{\alpha}\beta\tilde{\varphi}_{3}(g,\eta)\\
& \\
\bar{\alpha}\dot{\alpha}+\beta\overset{\cdot}{\bar{\beta}} &  =-\frac{1}%
{2}\left(  \alpha\beta-\bar{\alpha}\bar{\beta}\right)  \tilde{\varphi}%
_{1}(g,\eta)-\frac{1}{2}i\left(  \bar{\alpha}\bar{\beta}+\alpha\beta\right)
\tilde{\varphi}_{2}(g,\eta)
\end{align*}
where we denoted $\tilde{\varphi}(g,\eta)=\sum_{i=1}^{3}\tilde{\varphi}%
_{i}(g,\eta)X_{i}$. This system of equation can be solved for two components
of $\eta$, for instance $\eta_{1}$ and $\eta_{2}$, as a function of the
velocities and $\eta_{3}$,
\[%
\begin{array}
[c]{l}%
\eta_{1}=i\dfrac{\left(  \alpha\bar{\beta}-\beta\bar{\alpha}\right)  }%
{2\beta\bar{\alpha}\left\vert \beta\right\vert ^{2}}\left(  \bar{\alpha}%
\dot{\beta}-\beta\overset{\cdot}{\bar{\alpha}}\right)  \\
\qquad-i\dfrac{\left(  \left\vert \beta\right\vert ^{4}+\left\vert
\beta\right\vert ^{2}+\beta^{2}\bar{\alpha}^{2}\right)  }{2\beta\bar{\alpha
}\left\vert \beta\right\vert ^{4}}\left(  \bar{\alpha}\dot{\alpha}%
+\beta\overset{\cdot}{\bar{\beta}}\right)  -\dfrac{\left(  \left\vert
\alpha\right\vert ^{2}\left\vert \beta\right\vert ^{2}+\bar{\alpha}^{2}%
\beta^{2}\right)  }{2\beta\bar{\alpha}\left\vert \beta\right\vert ^{2}}%
\eta_{3}\\
\\
\eta_{2}=\dfrac{\left(  \alpha\bar{\beta}+\beta\bar{\alpha}\right)  }%
{2\beta\bar{\alpha}\left\vert \beta\right\vert ^{2}}\left(  \bar{\alpha}%
\dot{\beta}-\beta\overset{\cdot}{\bar{\alpha}}\right)  \\
\qquad-\dfrac{\left(  \left\vert \beta\right\vert ^{4}+\left\vert
\beta\right\vert ^{2}-\beta^{2}\bar{\alpha}^{2}\right)  }{2\beta\bar{\alpha
}\left\vert \beta\right\vert ^{4}}\left(  \bar{\alpha}\dot{\alpha}%
+\beta\overset{\cdot}{\bar{\beta}}\right)  +i\dfrac{\left(  \left\vert
\alpha\right\vert ^{2}\left\vert \beta\right\vert ^{2}-\beta^{2}\bar{\alpha
}^{2}\right)  }{2\beta\bar{\alpha}\left\vert \beta\right\vert ^{2}}\eta_{3}%
\end{array}
\]
Then, the lagrangian function is defined as
\[
L_{SU(2)}\left(  g,\dot{g}\right)  =\left\langle \eta,g^{-1}\dot
{g}\right\rangle -\mathcal{H}(g,\eta)
\]
where $\hat{\pi}$ was given in $\left(  \ref{dres orb annhilator}\right)  $.
Replacing $\eta_{1},\eta_{2}$ by the above relations, we obtain the
Lagrangian
\begin{equation}
L_{SU(2)}\left(  g,\dot{g}\right)  =\frac{1}{2}\left(  g^{-1}\dot{g}%
,g^{-1}\dot{g}\right)  _{\mathfrak{su}_{2}}+\eta_{3}\left\langle \hat{\pi
},g^{-1}\dot{g}\right\rangle \label{lagr dress inv syst}%
\end{equation}
where we have introduced the metric in trivialized tangent bundle
$SU(2)\times\mathfrak{su}_{2}$ given by%
\[
\left(  g^{-1}\dot{g},g^{-1}\dot{g}\right)  _{\mathfrak{su}_{2}}=-\frac
{1}{8\left\vert \beta\right\vert ^{2}}\kappa\left(  g^{-1}\dot{g},g^{-1}%
\dot{g}\right)
\]
Observe that $\eta_{3}$ appears as a Lagrange multiplier realizing the
constraint
\[
\left\langle \hat{\pi},g^{-1}\dot{g}\right\rangle \equiv\dfrac{i}{2\left\vert
\beta\right\vert ^{2}}\left(  \bar{\beta}\dot{\beta}-\beta\overset{\cdot}%
{\bar{\beta}}\right)  =0
\]
which in terms of the Euler angles reduces to%

\[
\dot{\psi}-\dot{\phi}=0
\]
showing that the dynamics is naturally restricted on dressing orbit, as expected.

By introducing $\mathrm{A}\left(  g\right)  :=8\left\vert \beta\right\vert
^{2}\eta_{3}\hat{\kappa}_{\mathfrak{su}_{2}}^{-1}\left(  \hat{\pi}\right)
\in\mathfrak{su}_{2}$, and after handling the quadratic form, we may write the
Lagrangian as
\[
L_{SU(2)}\left(  g,\dot{g}\right)  =\dfrac{1}{2}\left(  g^{-1}\dot
{g}-\mathrm{A}\left(  g\right)  ,g^{-1}\dot{g}-\mathrm{A}\left(  g\right)
\right)  _{\mathfrak{su}_{2}}-\dfrac{1}{2}\left(  \mathrm{A}\left(  g\right)
,\mathrm{A}\left(  g\right)  \right)  _{\mathfrak{su}_{2}}%
\]
that describe the dynamics of a particle moving on the group manifold $S^{3}$
under the action of non Abelian potential vector potential $\mathrm{A,}$ which
confines its movement to the $S^{2}$ sphere determined by the constraint
$\arg\beta=cte$.

\section{Conclusions}

We have shown how the theory of integrable systems, in particular AKS theory,
can be used in its full scope to solve effectively the systems involved in a
Poisson Lie T-duality scheme. In doing so, we have also introduced a variant
for the hamiltonian Poisson Lie T-duality scheme, by using as a central object
of the scheme a coadjoint orbit of one of the Iwasawa factors. This fact
enhances the applicability of the PL duality, including a wider class of
systems, finite or infinite dimensional, on which the techniques of integrable
systems can be used.

The explicit finite dimensional example $SL(2,\mathbb{C)=}$ $SU(2)\times B$
exhibits a detailed description of the way in which this duality works,
constructing explicitly the solutions for all the involved systems from the
factorization of the solution curve of an almost trivial system on
$\mathfrak{sl}_{2}^{\ast}$. This curve $\tilde{b}\left(  t\right)  \subset B$
gives rise to the solution curves in each case through the corresponding
actions. As an alternative way for using the scheme, the solutions would be
obtained retrieving the curve $\tilde{b}\left(  t\right)  \subset B$ from the
well known solution of the Toda system on $\mathbb{R}^{2}$. It is worth to
remark that the election of the symmetry group defines the master integrable
system ruling the dynamics and it is realized in this example by the choice of
$B$ as the main symmetry, putting the Toda system in the center of the scheme,
or in the loop group case of references \cite{CM} and \cite{CMZ}, where the
WZNW model appears on the double Lie group $LD=LG\times LG^{\ast}$. The
compatible dynamics was obtained from collective hamiltonian functions after
fixing a hamiltonian on the selected coadjoint orbit.

The systems in the equivalence class includes a kind of generalized top on the
group $B$ and a dressing invariant system on the group manifold $S^{3}$ which
suffers a reduction to the $S^{2}$ submanifold characterized by $\arg\beta$.
Dressing symmetry becomes relevant for the so called \emph{Poisson sigma
models}, so our example may serve as a laboratory for understanding issues
related to the reduced space of systems with this kind of symmetry. 

From the Lagrangian point of view, the PL T-duality transformation relates a
constrained systems on the compact configuration space $SU(2)$ with a system
on the non compact space $B$, by a rather non trivial transformation. A
remarkable fact is that these non-linear systems arise from kinetic
lagrangians, that means, bilinear forms on the corresponding tangent bundles.
In the $SU(2)$ case the bilinear form amounts to be metric, while in $B$ case,
because a solvable Lie group lacks of an Ad-invariant bilinear form on it, the
bilinear form is inherited from $\mathfrak{sl}_{2}$ through a linear operator
$\mathbb{K}$. This relation between two different \emph{target geometries}
relies on the dynamical equivalence of the reduced hamiltonian systems and the
coadjoint orbit.  In both cases, the structure of the reduced phase spaces
were explicitly determined, and the PL T-duality equivalence between the
lagrangian system $\left(  \ref{Toda Lag}\right)  $, $\left(
\ref{full lagrangian for B II}\right)  $ and the $\left(
\ref{lagr dress inv syst}\right)  $ was established.

Most of the theory of integrable systems applied in this work can be used,
with some cares, in the infinite dimensional case (loop groups). In fact, the
references \cite{AKS}, \cite{RSTS 1} and \cite{RSTS 2} deal with Kac-Moody
algebras and infinite dimensional integrable systems like KdV and others, so
we expect they can be applied in the natural setting of Poisson Lie T-duality,
namely the loop groups case and T-dualizable sigma models.

\section{Acknowledgments}

The author thanks to CONICET for financial support.

%\section{References}


\begin{thebibliography}{99}                                                                                               %


\bibitem {KS-1}C. Klimcik, P. Severa, \emph{Poisson-Lie T-duality and loop
groups of Drinfeld doubles}, Phys. Lett. B 351, 455-462 (1995),
hep-th/9512040; \emph{Dual non-Abelian duality and the Drinfeld double}, Phys.
Lett. B 372, 65-71 (1996), hep-th/9502122; \emph{Poisson-Lie T-duality}; C.
Klimcik, Nucl. Phys. Proc. Suppl. 46, 116-121 (1996), hep-th/9509095.

\bibitem {CM}A. Cabrera, H. Montani, \emph{Hamiltonian loop group actions and
T-Duality for group manifolds}, J. Geom. Phys. 56 (2006), 1116-1143; hep-th/0412289.

\bibitem {CMZ}A. Cabrera, H. Montani, M. Zuccalli, \emph{Poisson Lie T-Duality
and non trivial monodromies}, J. Geom. Phys. 59 (2009), 576-599; math-phys/0712.2259.

\bibitem {AKS}M. Adler, P. van Moerbeke, \emph{Completely integrable systems,
Euclidean Lie algebras and curves}, Adv. Math. 38 (1980), 267-317; B. Kostant,
\emph{The solution to a generalized Toda lattice and representation theory},
Adv. Math. 34 (1979), 195-338; W. Symes, \emph{Systems of Toda type, inverse
spectral problem and representation theory}, Inv. Math 159 (1980), 13-51.

\bibitem {Hel}S. Helgason, \emph{Differential geometry, Lie groups, and
symmetric spaces}, Academic Press, New York, 1978.

\bibitem {Lu-We}J.-H. Lu, A. Weinstein, J. Diff. Geom. 31, 501-526 (1990).

\bibitem {ChPr}V. Chari, A. Pressley, \emph{A Guide to Quantum Groups},
Cambridge University Press, 1994.

\bibitem {Abr-Mars}M. Abraham, J. Marsden, \emph{Foundations of Mechanics},
Massachusetts: Benjamin/Cummings, Reading, 2nd. ed., (1978).

\bibitem {STS}M.A. Semenov-Tian-Shansky, \emph{Dressing transformations and
Poisson group actions, }Publ. RIMS, Kyoto Univ. 21 (1985), 1237-1260.

\bibitem {Mars-Wein}J. E. Marsden, A. Weinstein, \emph{Reduction of symplectic
manifolds with symmetry}, Rep. Math. Phys. 5, 121-131 (1974).

\bibitem {Guill-Sten}V. Guillemin, S. Sternberg, \emph{Symplectic techniques
in physics}, Cambridge, Cambridge Univ. Press, 1984.

\bibitem {KKS}D. Kazhdan, B. Kostant , S. Sternberg, \emph{Hamiltonian group
actions and dynamical system of Calogero type}, Commun. Pure Appl. Math. 31,
481-508 (1978).

\bibitem {holm-marsden}D. Holm, J E Marsden, \emph{The rotor and the
pendulum}, in Symplectic Geometry and Mathematical Physics, P. Donato, C.
Duval, J. Elhadad, G. M. Tuynman, ed., Prog. in Math. Vol. 99, Birkhauser:
Boston, 1991, pp. 189--203.

\bibitem {RSTS 1}Reyman, A. G. \& Semenov-Tian-Shansky, M. A., \emph{Reduction
of hamiltonian systems, affine Lie algebras, and Lax equations} \emph{I},
Invent. Math. 54, 81-100 (1979)

\bibitem {RSTS 2}Reyman, A. G. \& Semenov-Tian-Shansky, M. A., \emph{Reduction
of hamiltonian systems, affine Lie algebras, and Lax equations II}, Invent.
Math. 63, 423-32 (1981)
\end{thebibliography}
\end{document}